\pdfoutput=1
\documentclass[aps,pre, twocolumn, groupedaddress]{revtex4-1} 
\usepackage{lipsum}
\usepackage{mathtools}
\usepackage{graphicx}

\usepackage{relsize}
\usepackage{amsmath}    % need for subequations
\usepackage{amssymb}
\usepackage{bm}
\usepackage{hyperref}
\usepackage{latexsym}
\usepackage{verbatim}
\usepackage{color}
\usepackage[caption=false]{subfig}

\def\beq{\begin{equation}}
\def\eeq{\end{equation}}

\def\beq{\begin{equation}}                          
\def\eeq{\end{equation}}                          
\def\bea{\begin{eqnarray}}                          
\def\eea{\end{eqnarray}}

\DeclareRobustCommand{\uvec}[1]{{%
  \ifcsname uvec#1\endcsname
     \csname uvec#1\endcsname
   \else
    \bm{\hat{\mathbf{#1}}}%
   \fi
}}

\preprint{}

\begin{document}

%%%%%%%%%%%%%%%%%%%%%%%%%%%%%%%%%%%%%%%%%%%%%%%%%%%
%                               TITLE & ABSTRACT
%%%%%%%%%%%%%%%%%%%%%%%%%%%%%%%%%%%%%%%%%%%%%%%%%%%
\title{Ordering kinetics and steady states of XY-model with ferromagnetic and nematic interaction}
\author{Partha Sarathi Mondal}
\email{parthasarathimondal.rs.phy21@itbhu.ac.in}
\affiliation{Indian Institute of Technology (BHU) Varanasi, India 221005}
\author{Pawan Kumar Mishra}
\email{pawankumarmishra.rs.phy19@itbhu.ac.in}
\affiliation{Indian Institute of Technology (BHU) Varanasi, India 221005}
\author{Shradha Mishra}
\email[]{smishra.phy@itbhu.ac.in}
\affiliation{Indian Institute of Technology (BHU) Varanasi, India 221005}
\date{\today}
\begin{abstract}
{Previous studies on the generalized XY model have concentrated on the equilibrium phase diagram and the equilibrium nature of distinct phases under varying parameter conditions. We direct our attention towards examining the system's evolution towards equilibrium states across different parameter values, specifically by varying the relative strengths of ferromagnetic and nematic interactions. We study the kinetics of the system, using the temporal annihilation  of defects at varying temperatures and its impact on the coarsening behavior of the system. For both pure polar and pure nematic systems, we observe temperature-dependent decay of the exponent, leading to a decelerated growth of domains within the system. At parameter values where both ferromagnetic and nematic interactions are simultaneously present, we show a phase diagram highlighting three low-temperature phases—polar, nematic, and coexistence—alongside a high-temperature disordered phase. Our study provides valuable insights into the complex interplay of interactions, offering a comprehensive understanding of the system's behavior during its evolution towards equilibrium. 
}
\end{abstract}
\maketitle
\section{Introduction}
The two-dimensional (2D) XY model has attracted substantial attention due to its diverse applications in magnetic systems, along with its relevance to quantum liquids and superconductors. The seminal work of N.D. Mermin and H. Wagner \cite{mermin1967absence} demonstrated that in the 2D XY model, the sustainability of long-range order is inherently limited. Even a very small thermal fluctuations have the profound capability to disrupt the long-range order within the system. Instead, the system exhibits Quasi Long Range Order (QLRO) and experiences a Berezinskii-Kosterlitz-Thouless (BKT) transition, leading to a disordered state through the unbinding of defect pairs. Following the works of Berezinskii, Kosterlitz \cite{BKT} and Thouless \cite{berezinskii1971destruction}, this model has been studied extensively \cite{LPTGF,jensen2003kosterlitz,packard2013introduction,Yurke1993,Lasher1972,Jonathan2018,vanderstraeten2019approaching,mila1993first,RichtMon}.\\
The extension of the XY model to incorporate an additional nematic interaction, alongside the ferromagnetic interaction, was introduced by Korshunov \cite{korshunov1985possible}, Lee and Grinstein \cite{lee1985strings}. Subsequently, this model has since become a focal point of extensive research due to its significance in liquid crystals \cite{lee1985strings,domany1984first,pang1992string} and superfluid $^3$He films \cite{korshunov1985possible,bhaseen2012discrete}. In these systems, an additional nematic interaction with periodicity $\frac{2\pi}{q}$ coexists with the conventional magnetic interaction featuring periodicity $2\pi$. This gives rise to the formation of vortices characterized by a topological charge of $\frac{1}{q}$ alongside conventional integer defects. For the case of $q=2$, the nematic interaction results in an equal probability of parallel and antiparallel alignment of each spin with its neighbors, thereby competing with the parallel alignment tendency due to the ferromagnetic interaction. The interplay between these interactions gives rise to intriguing properties of the system, as elucidated in previous studies \cite{D_B_Carpenter_1989,Benakli_1997,Qin2009,Milan2016,Park2008,Maccari,Poderoso2011,Lee1985,QI2013127,shi2011boson}. These investigations delved into various aspects of the model, such as the phase diagram \cite{dian2011spin}, the nature of order in the system \cite{vzukovivc2019x}, the characteristics of the phase transition \cite{canova2014kosterlitz,serna2017deconfinement,mishra2022active}, the behavior of the correlation length, stiffness jump across the transition \cite{hubscher2013stiffness}, finite-size effects \cite{nui2018correlation}, and the structure of defects \cite{kobayashi2020z}.\\
While equilibrium states have been extensively explored, the kinetics of the system's transition to equilibrium from a random initial condition remains largely unexplored. As the system evolves towards the equilibrium state, the physics of integer and half-integer defects become pivotal in determining the system's fate. Therefore, understanding the kinetics of the system in terms of statistics of defects is crucial for developing a comprehensive understanding of its behavior. In this context, a fundamental question arises: How does the relative strength of ferromagnetic and nematic interactions influence the  defect annihilation and, consequently, the coarsening kinetics of the system?

This study focuses on elucidating the statistics of defects with varying temperature and relative strengths of the two types of interactions, influencing the overall coarsening kinetics of the system. Unlike previous works which predominantly explored the equilibrium phase diagram, in this study, we explore the phase diagram based on the kinetics of the system. The investigation encompasses extreme cases—pure ferromagnetic and pure nematic—as well as the intermediate regime where both interactions coexist (mixed system) for temperature $T<T_{BKT}$, where $T_{BKT}$ is BKT transition temperature.

To conclude this introduction, we provide a concise summary of our key results: (i) In the pure ferromagnetic/nematic limit of the model, the annihilation of the defects get notably slower with the increase of temperature. (ii) in the intermediate regime (mixed system), when both ferromagnetic and nematic interactions are present simultaneously, the system exists in distinct phases based on different combinations of parameters. Particularly, when the strengths of ferromagnetic and  nematic interactions are comparable, we identify a state where both types of defects coexist within the system.\\
The forthcoming organization of this paper is structured as follows: Section \ref{secII} provides an elaborate discussion of the particulars of our model. Section \ref{secIII} delves into the results, where we first examine the dynamic of defects in a Pure Polar System in Section \ref{secIIIA1}, followed by an analysis of the Pure Nematic System in Section \ref{secIIIA2}. We then extend our analysis to Mixed Systems in Section \ref{secIIIA3}. Next, we illustrate the Phase Diagram in Section \ref{secIIID}. Finally, our study is encapsulated in Section \ref{secdis}, where we offer a comprehensive summary of our findings.

\twocolumngrid
\begin{figure*} [hbt]
\centering
\subfloat[]{\includegraphics[width=0.495\textwidth]{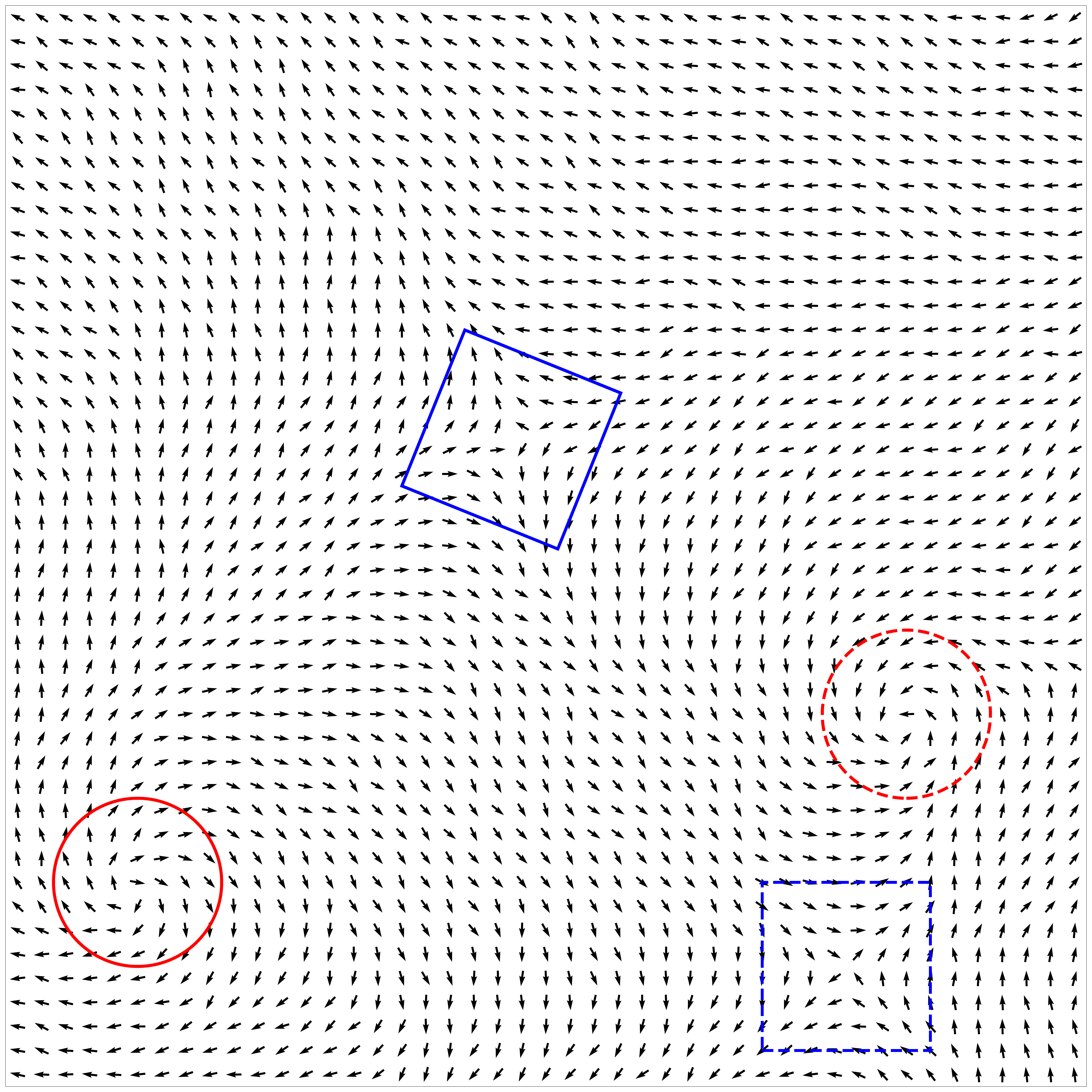}}
~
\subfloat[]{\includegraphics[width=0.495\textwidth]{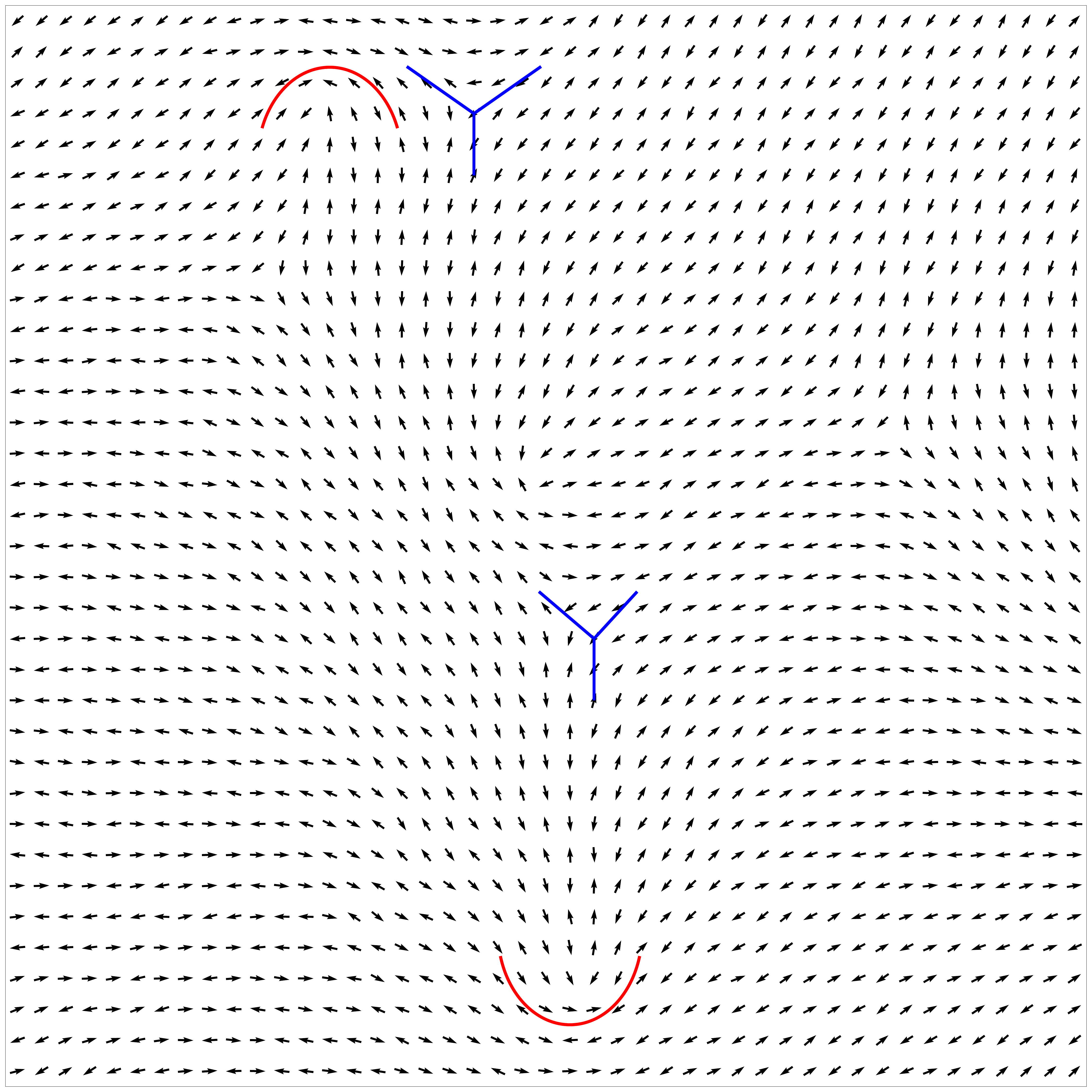}} 
	\caption{(color online) The figure illustrates defect configurations in the system with pure ferromagnetic interaction, subplot (a), and pure nematic interaction, subplot (b). The black quivers represent the spins at each lattice point. (a) Shows the $\pm 1$ defects with red circle and blue square, respectively. Left and right hand vortices with same winding number are marked by solid and dashed symbol, respectively. (b) shows $\pm \frac{1}{2}$ defect configurations marked with  red arcs, and blue `Y'-shaped structures, respectively. A part of the system of size $L=128$ is shown for better resolution.}
\label{fig:1}
\end{figure*}

\twocolumngrid
\begin{figure*}[ht]
\centering
\subfloat[]{\includegraphics[width=0.45\textwidth]{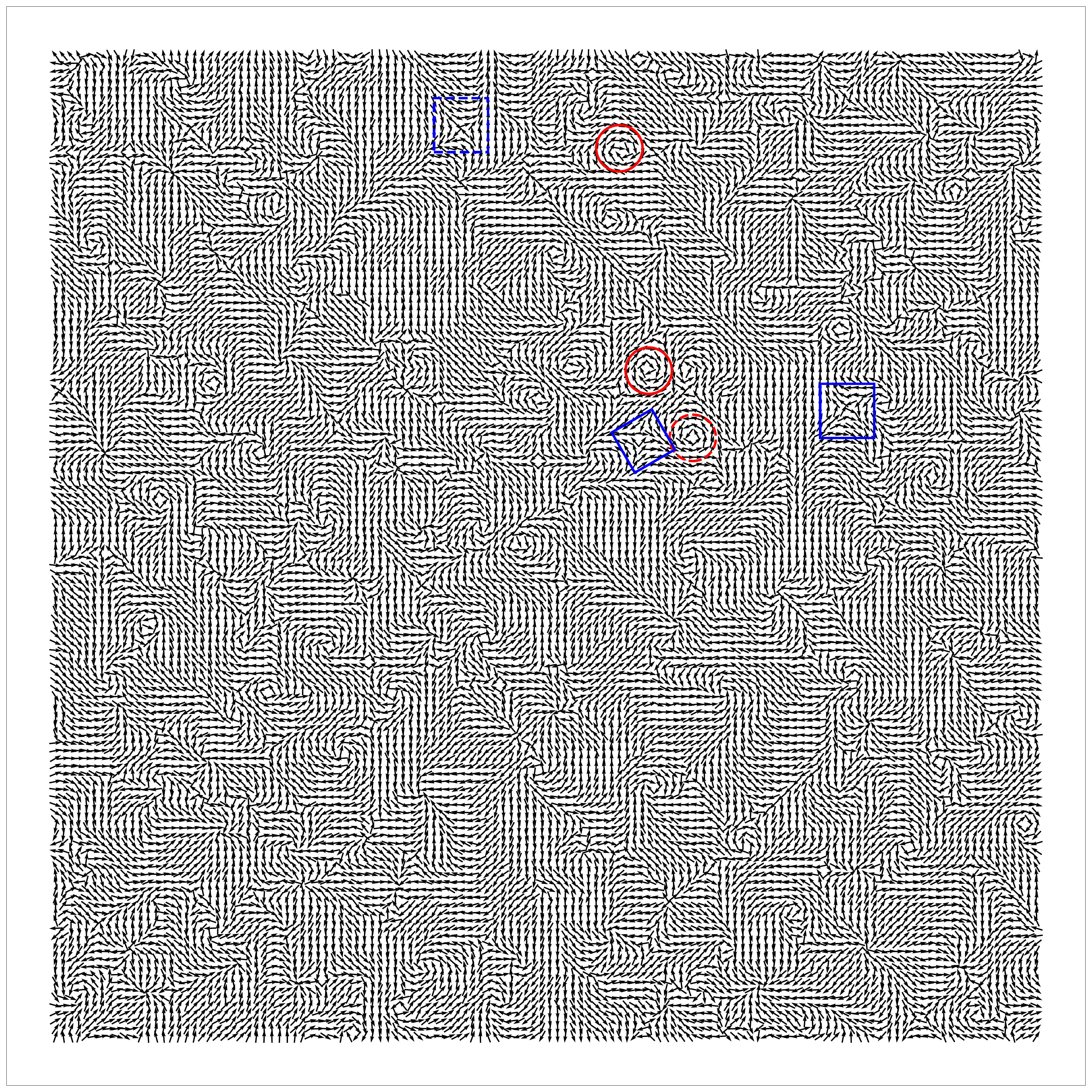}}
~
\subfloat[]{\includegraphics[width=0.45\textwidth]{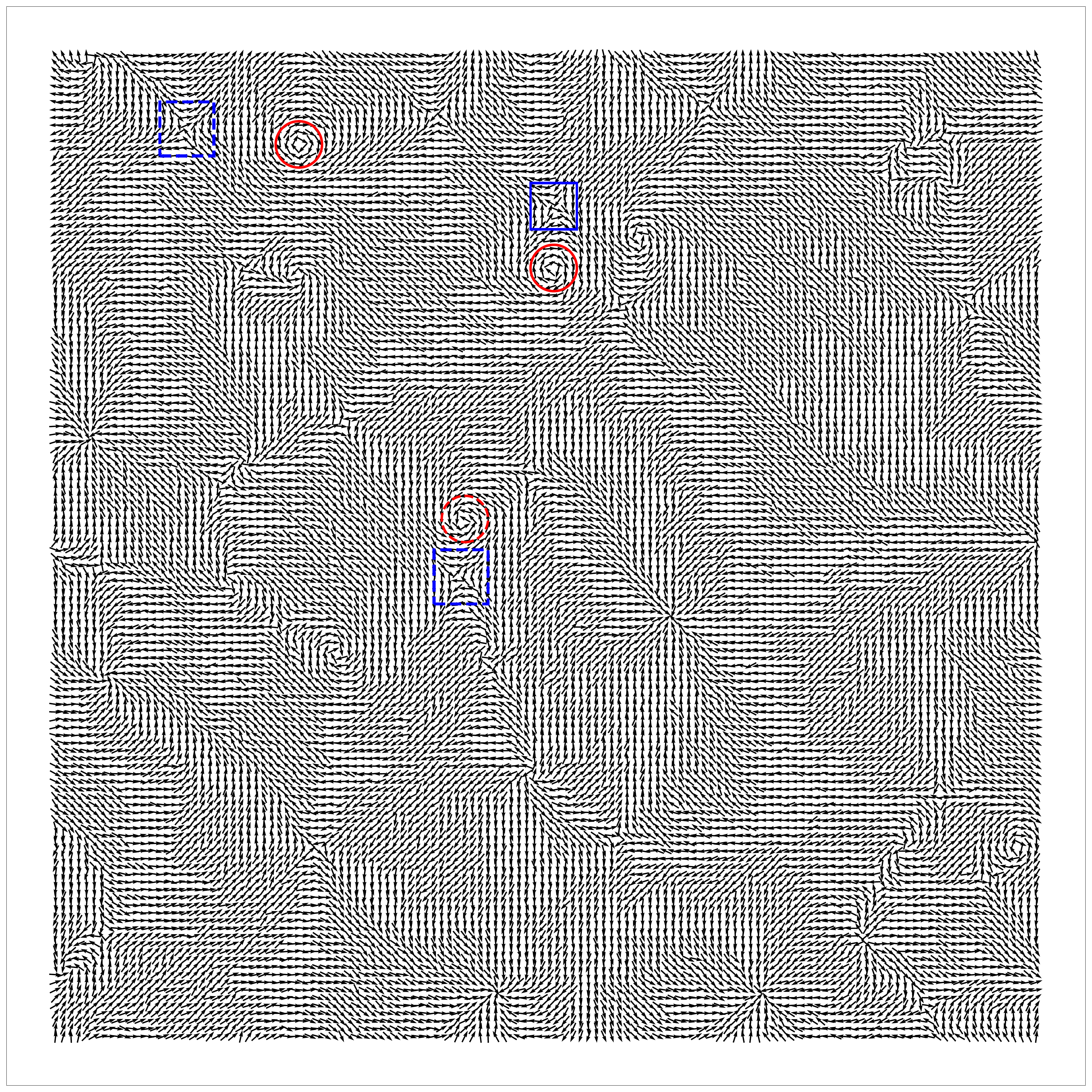}}

\subfloat[]{\includegraphics[width=0.45\textwidth]{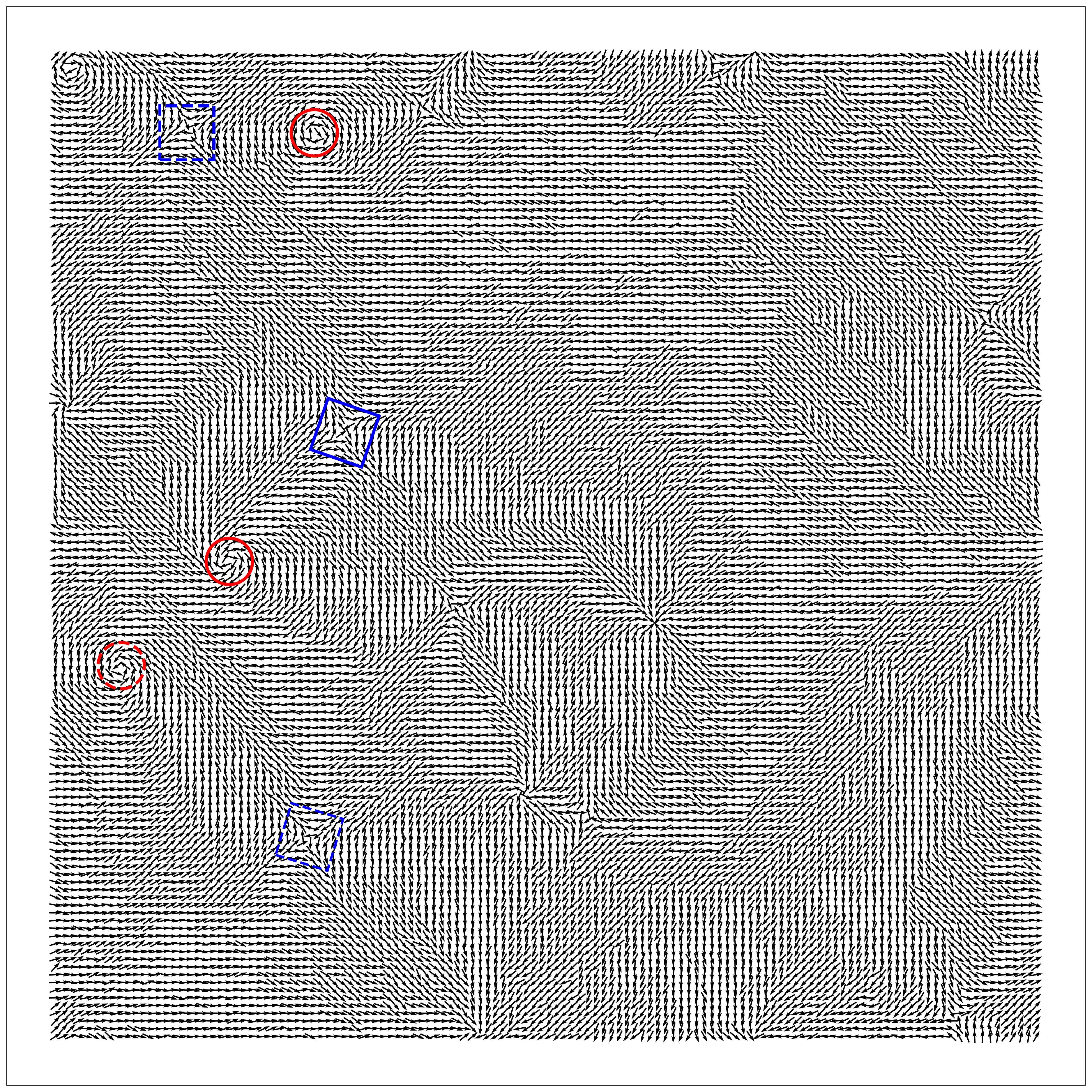}}
~
\subfloat[]{\includegraphics[width=0.45\textwidth]{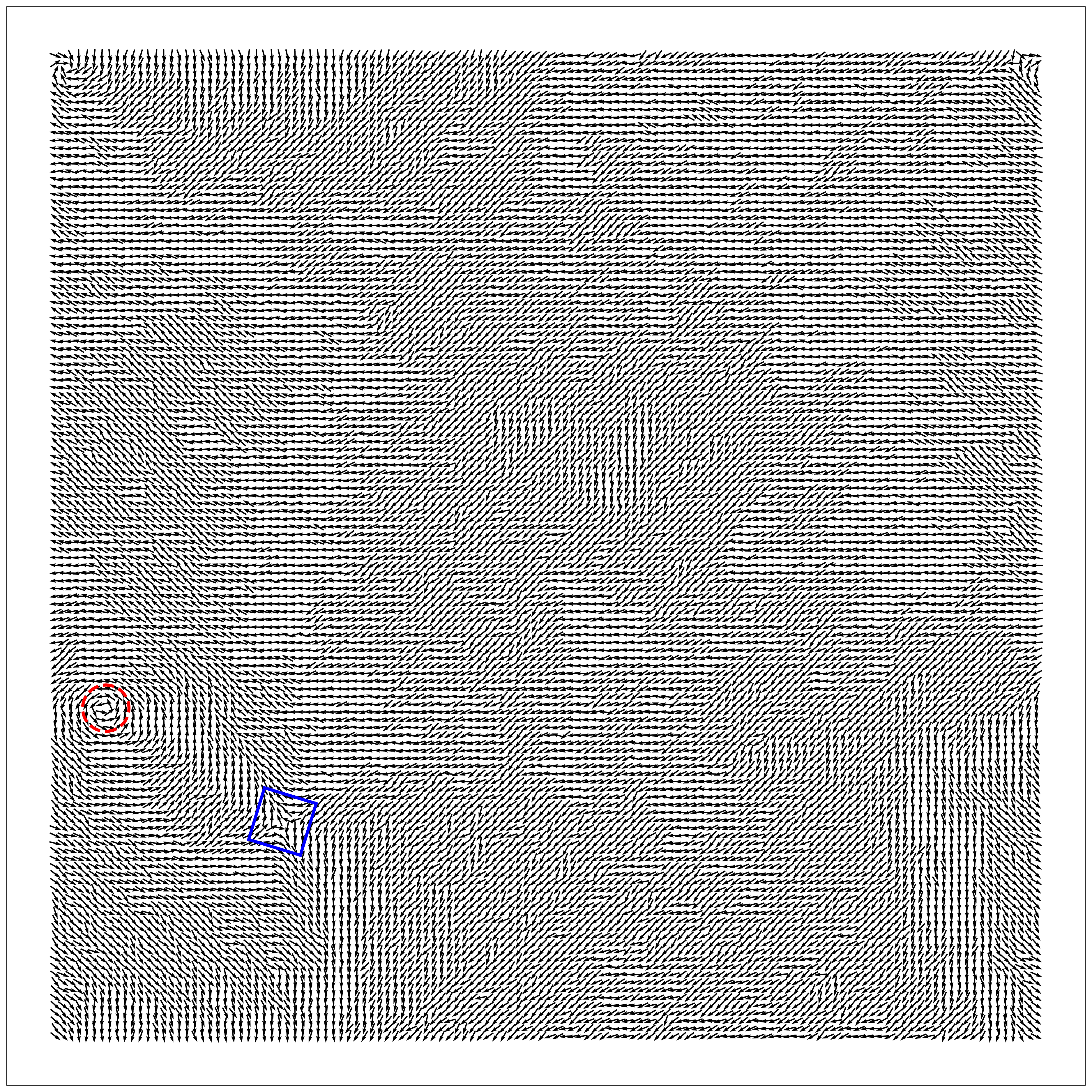}}

\caption{(color online) The figure provides a visual representation of the evolving Pure Polar System, depicted through a sequence of snapshots. Subplots (a)-(d) exhibit snapshots of the system at distinct time instances: 200, 1000, 2000, and 6000, respectively. In each subplot, the black quivers represent the spins at each lattice sites. Subplots (a)-(d) highlight certain $+1$ and $-1$ defects, marked by red circles and blue squares, respectively. Left and right vortices of same charge are marked by solid and dashed symbols; System Size L = $128$ and $\frac{T}{T_{BKT}} = 0.3$.}
\label{fig:2}
\end{figure*}

\begin{figure} [hbt]
{\includegraphics[width=1.0 \linewidth]{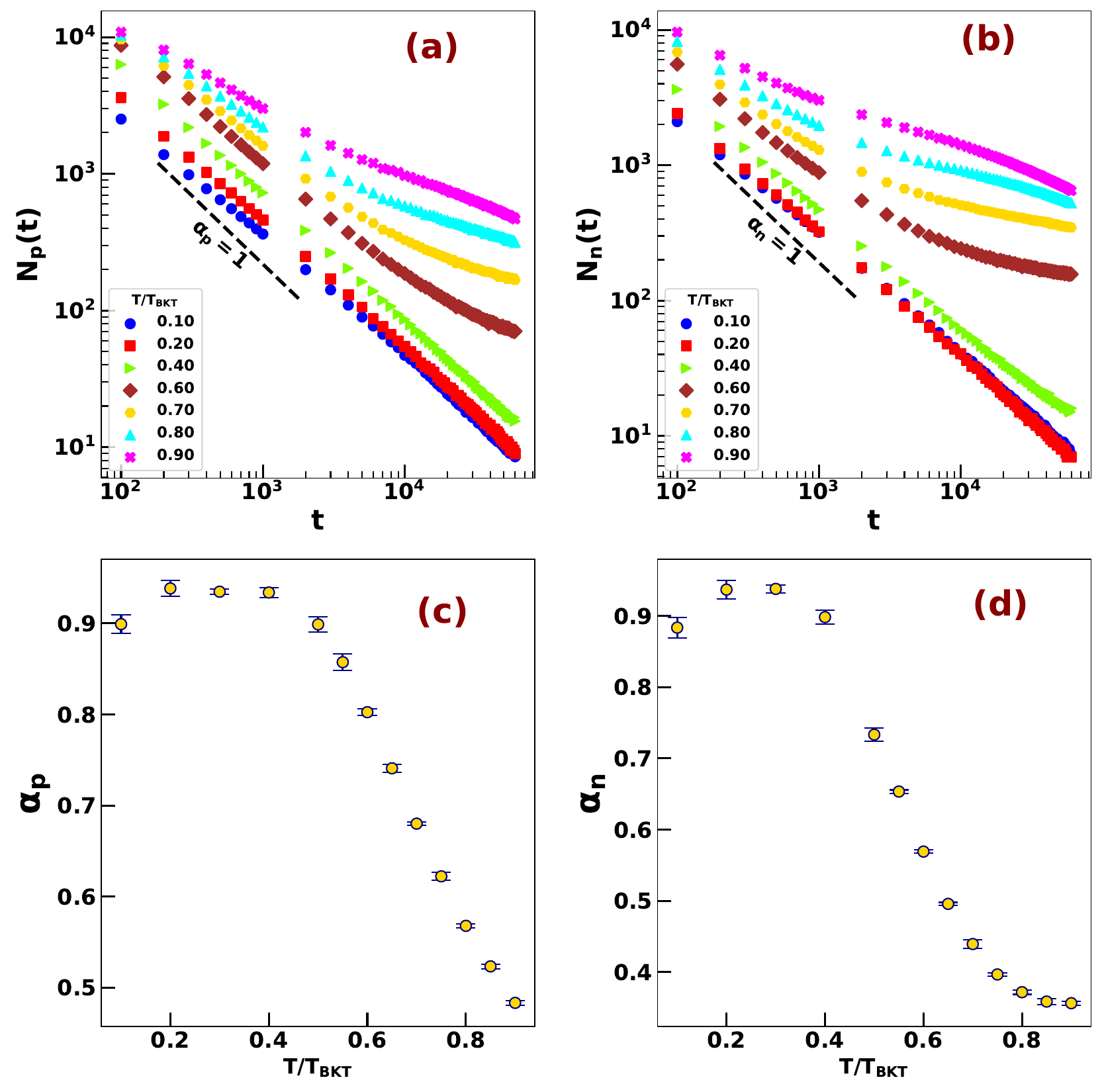}}
	\caption{(color online) The figure demonstrates the temporal evolution of the system for both the Pure Polar and Pure Nematic cases. The left panel (subplots (a) and (c)) illustrates the Pure Polar case, while the right panel (subplots (b) and (d)) depicts the Pure Nematic case. Subplots (a) and (b) showcase the time-dependent decay of defects, $N_p(t)$ and $N_n(t)$,  for pure polar and pure nematic systems, respectively, at different temperatures on a log-log scale. In both subplots, the black dashed line indicates a power-law decay with an exponent of 1 in the region where the slope is calculated. Subplots (c) and (d) present the variation of the exponent with temperature for the corresponding systems, with each data point accompanied by an error bar representing the uncertainty in the exponent values. The system size is fixed at $L=$ 256.}
\label{fig:3}
\end{figure}

\twocolumngrid
\begin{figure*}
\centering
\subfloat[]{\includegraphics[width=0.48\textwidth]{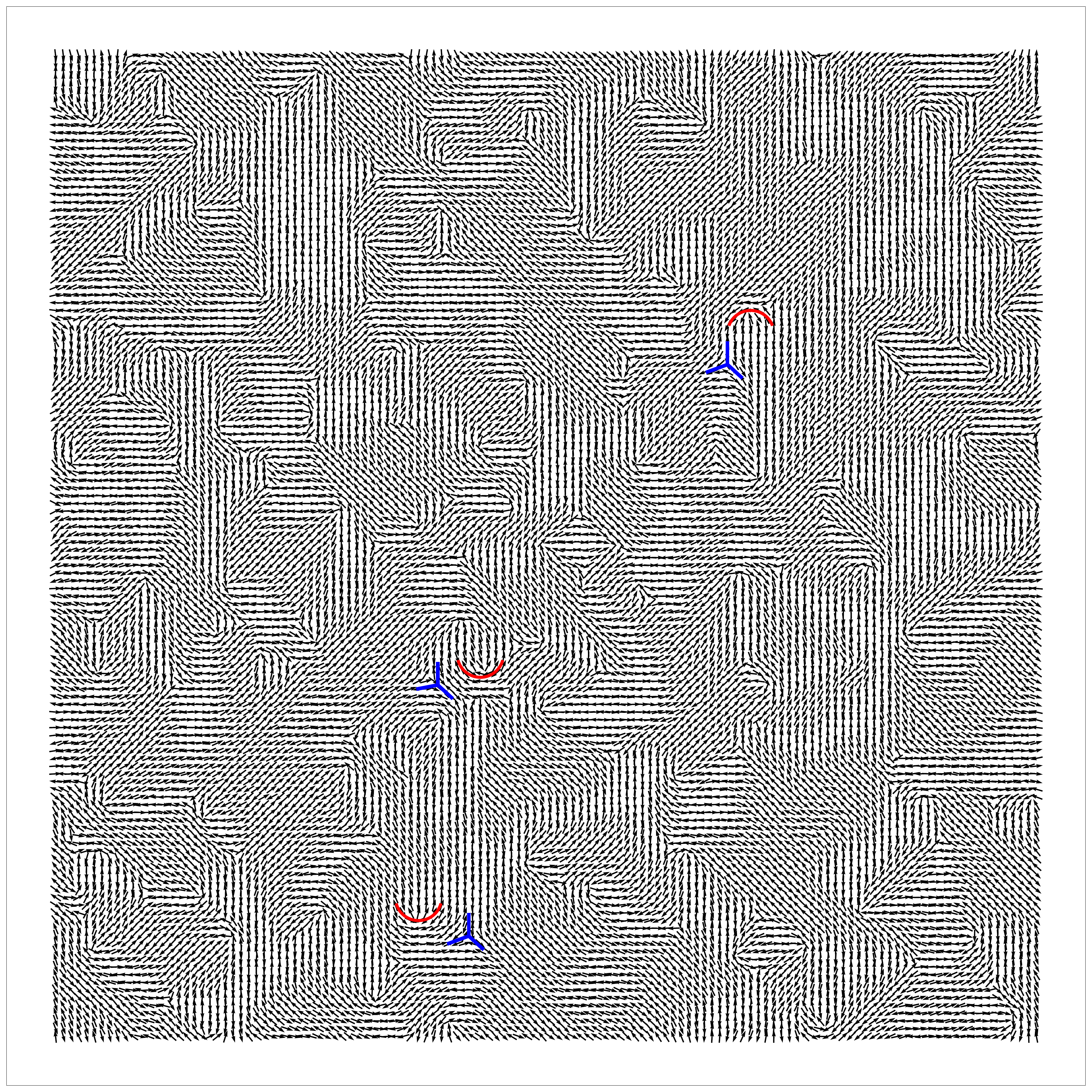}}
~
\subfloat[]{\includegraphics[width=0.48\textwidth]{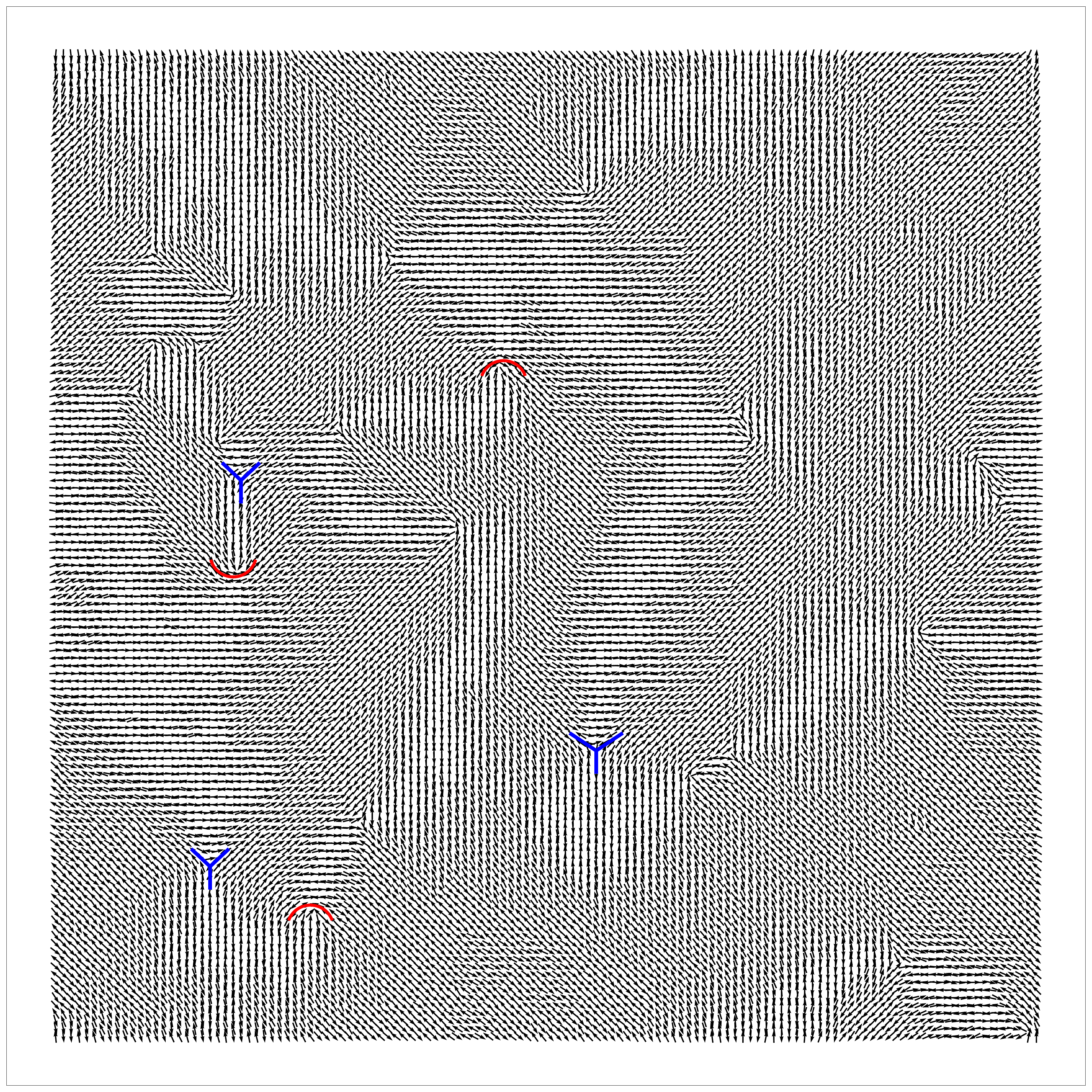}}

\subfloat[]{\includegraphics[width=0.48\textwidth]{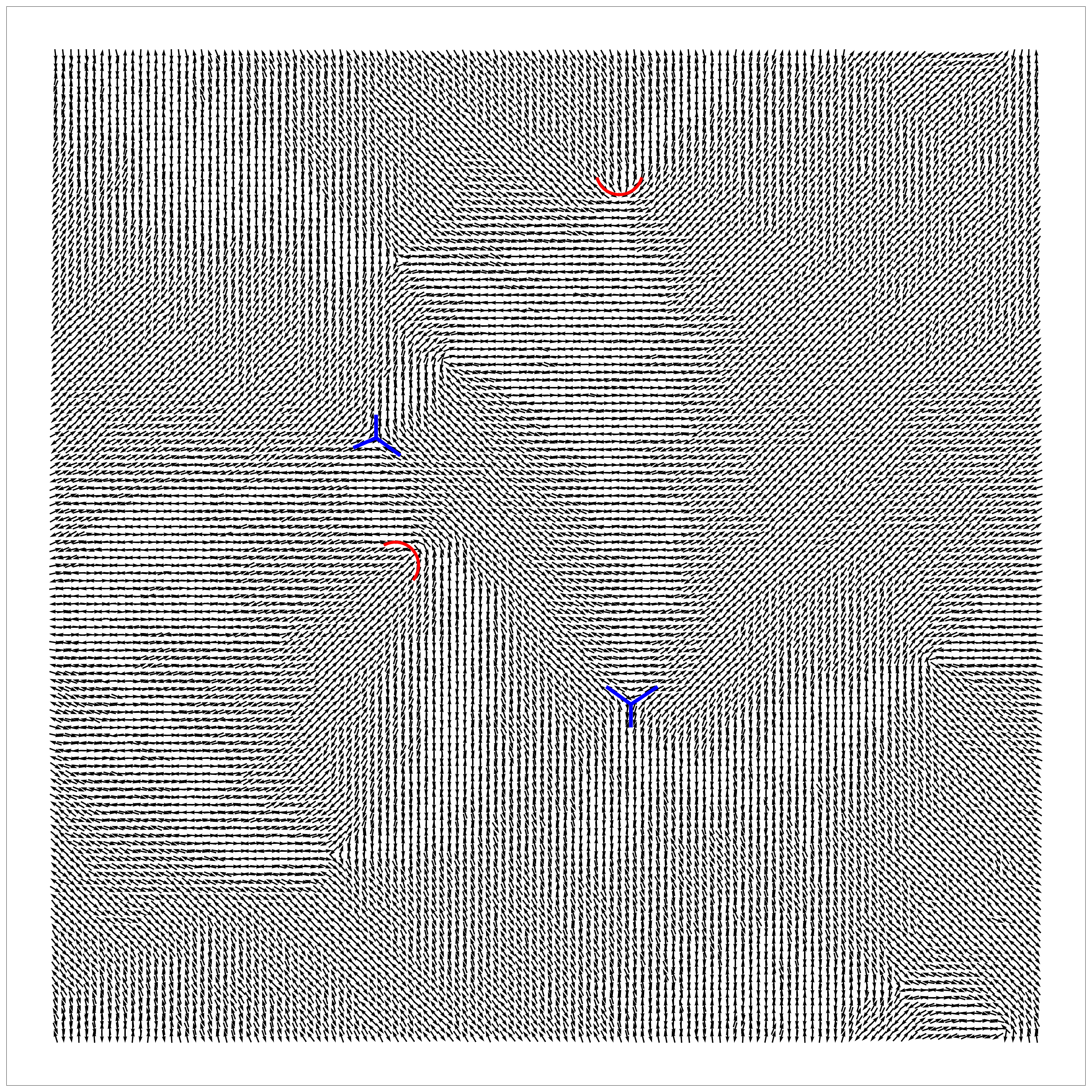}}
~
\subfloat[]{\includegraphics[width=0.48\textwidth]{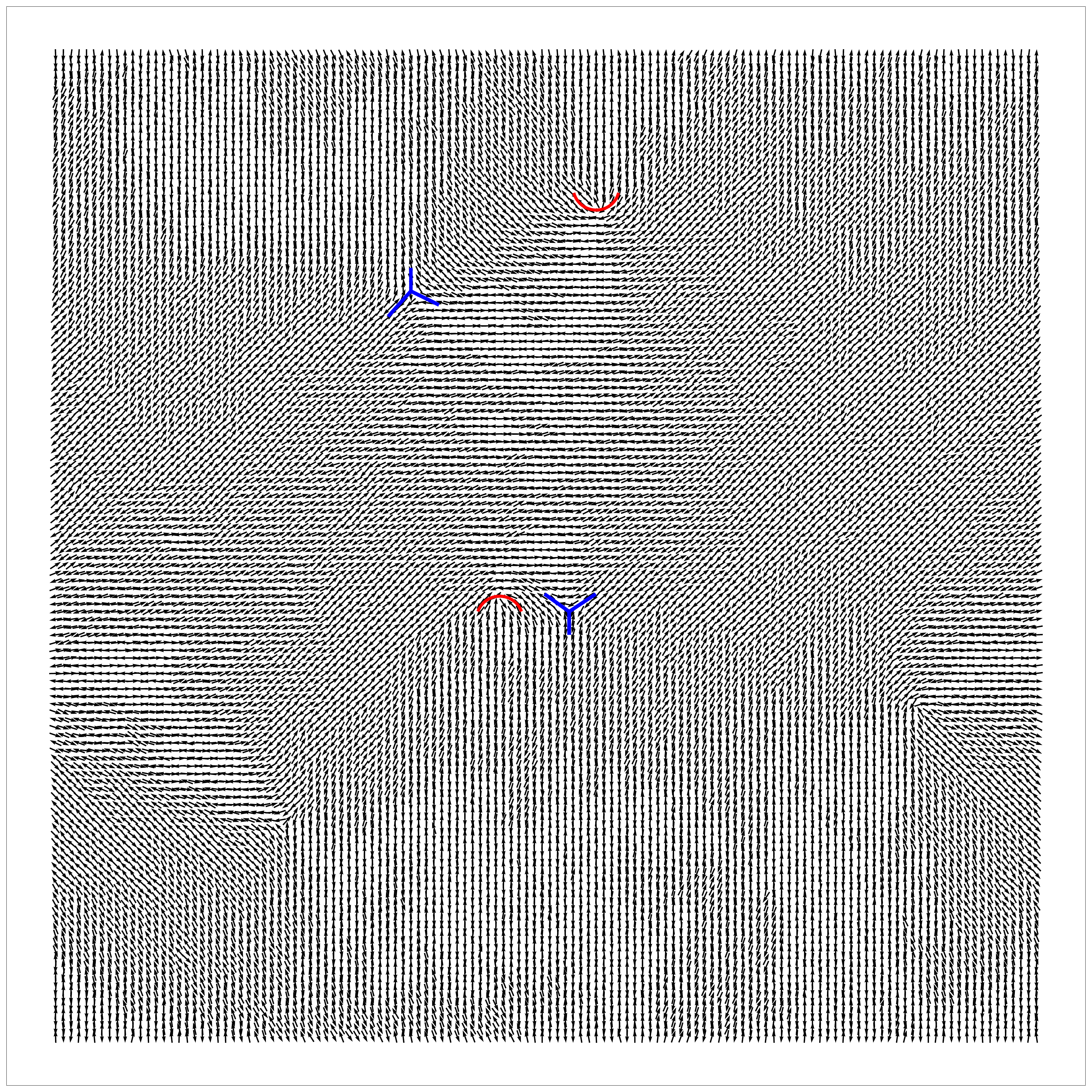}}

\caption{The presented figure captures the dynamic evolution of a Pure Nematic System, depicted sequentially through a set of snapshots. Subplots (a)-(d) present snapshots of the system taken at distinct time instances: 100, 1000, 2000, and 6000, respectively. In each subplot, the spins are represented by the black quivers. Notably, subplots (a)-(d) prominently showcase distinct $+\frac{1}{2}$ and $-\frac{1}{2}$ defects, which are marked by red arcs and blue `Y'-shaped structures, respectively. The remaining parameters are consistent with fig.\ref{fig:2}.}
\label{fig:4}
\end{figure*}

\section{Methodology}\label{secII}
\subsection{Model}\label{secIIA}
Our model considers a square lattice with spins arranged at unit lattice spacing. Each spin is represented by an angle $\theta$ ranging from 0 to $2\pi$. We express the spin as a vector $\vec S \equiv (\cos\theta, \sin\theta)$, where the magnitude is normalized to unity, i.e., $|\vec S| = 1$.

Here, the Hamiltonian of the classical XY model incorporates both ferromagnetic and nematic interactions, leading to the following modified form:
\begin{equation}
  H = -\Delta \sum_{<ij>} \cos(\theta_{ij}) - (1-\Delta) \sum_{<ij>} \cos(2\theta_{ij})
  \label{eq(1)}
\end{equation}
Here, $\theta_{ij}$ denotes the angle between spins $i$ and $j$, defined as $\theta_{ij} = \theta_i - \theta_j$. The summation is taken over pairs of nearest neighbors, indicated by $<ij>$. The parameters $\Delta$ and $(1-\Delta)$ correspond to the strengths of ferromagnetic and nematic interactions, respectively. The limiting cases $\Delta = 1$ and $\Delta = 0$ correspond to the pure ferromagnetic (polar) and pure nematic (apolar) cases, respectively. For the intermediate regime, $\Delta\in (0, 1)$, the interaction of each spin with it's neighbours has a ferromagnetic part and a nematic part with strengths given by $\Delta$ and $(1-\Delta)$, respectively. In the rest of this paper, we follow the following terminology: The spins are referred as polar and the state of the system is referred to as ``{\em Pure Polar}" when $\Delta = 1$, the spins are referred as apolar and the state of the system as ``{\em Pure Nematic}" when $\Delta = 0$, and ``{\em Mixed}" for $0 < \Delta < 1$. \\
While considering negative values of $\Delta$ can be interesting, we presently confine our observations to the range $\Delta \in (0,1)$ as physical systems corresponding to negative $\Delta$ values are presently unknown to us. Also in most of the previous studies,  the equilibrium properties  of the model are  explored for $\Delta \in (0, 1)$ limit \cite{D_B_Carpenter_1989,Benakli_1997,Qin2009,Milan2016,Park2008,Maccari,Poderoso2011,Lee1985,QI2013127,shi2011boson}. The parameter $\Delta <0$  also shows interesting behaviour as found in \cite{venditti2023phenomenological}. \\
The system comprises spins governed by the Hamiltonian specified in Eq. \ref{eq(1)}, arranged on a square lattice with dimensions $L \times L$ and subject to periodic boundary conditions (PBC) in both the $X$ and $Y$ directions. The evolution of the system is performed using the Metropolis-Monte Carlo algorithm. The study investigates the system's behavior by varying the temperature ($T < T_{BKT}$) and the parameter $\Delta$. For each $\Delta$ value, the system is initialized with spins randomly oriented, representing a high-temperature state ($>T_{BKT}$). The temperature $T$ is then fixed to the desired value ($<T_{BKT}$), leading to a sudden quench. The system subsequently evolves towards its equilibrium state at temperature $T$. The study encompasses an exploration of  the kinetic of  the system towards the equilibrium state when quenched from very high temperature as well as  the properties characterizing the resulting equilibrium state.\\
In our model, distances are measured in terms of the lattice spacing, and we set the Boltzmann constant $k_B$ to unity for consistency. The presented results are based on a $L=256$ system size unless stated otherwise. To ensure statistical accuracy, data are averaged over a minimum of $100$ independent realizations. Each realization involves simulating the system for $1.5 \times 10^4$ steps for a $L = 128$ system, or $3 \times 10^4$ steps for a $L = 256$ system, or $7 \times 10^4$ steps for a $L = 512$ system.\\
We also conduct simulations for system sizes, $L =$ $64$, $128$, $160$ and $200$ for the equilibrium phase diagram. For this purpose, we simulate the system for a total of $4.5 \times 10^5$ time steps, of which in the first $3.5 \times 10^5$ time steps, we allow the system to reach the corresponding equilibrium state and the rest $1 \times 10^5$ times steps are used for calculating various observable. For better statistics, we perform averaging over $10$-$15$ ensembles depending on the system size.

\subsection{Methods}\label{secIIB}
{\bf (a) Metropolis Montecarlo Algorithm (MCA):}  Our algorithm is outlined as follows: \\
\begin{enumerate}
    \item {We choose a spin at site $(i,j)$ and give a small random rotation to its direction, $\theta(i,j) = \theta(i,j) + \delta$, and calculate the change in energy $\Delta E$ by the rotation.}
    \item{The new orientation of the spin is accepted with probability, $P=min(1,e^{-\beta \Delta E})$, where $\beta = \frac{1}{T}$ denotes the inverse temperature.}
\end{enumerate}

 A single simulation step is considered complete when each spin in the system is updated once. We use the MCA to evolve the system with time and generate the configurations. Further, we use the configuration of the system at different times to calculate number of defects, correlation function etc. Similar method is previously used  to study the kinetics and scaling in different systems, including: Ising model \cite{paul2005domain}, $q$-state potts model \cite{sahni1983kinetics, kaski1985domain} and $XY-$model \cite{kumar2017ordering}.\\
 {\bf (b) Kinetics of the system using defect statistics:} In this study, we characterize the properties of the different phases  in terms of topological defects. First, we briefly describe the origin and importance of defects in a system with continuous symmetry.\\
Topological defects represent distinct distortions in the state of broken symmetry within a system. In its pursuit of equilibrium, the system seeks to minimize its free energy, hence favoring the ordered parallel or anti parallel alignment of spins in  ferromagnetic or nematic settings, respectively. However, the presence of defects introduces energy penalties. Remarkably, these defects possess topological stability and can only be eliminated through abrupt local changes in spin orientation. Consequently, achieving a defect-free configuration incurs a substantially higher energy cost. Thus, in two dimensions, the system exhibits a preference for configurations that include defects.\\
Topological defects feature a core where order is completely disrupted, while $\theta(x)$ exhibits slow variation in its distant surroundings. These defects are classified based on the winding number, denoted as $k$, which quantifies the change in $\theta(x)$ along a closed loop encircling the defect. The winding number can be positive or negative, as well as an integer or half-integer. Defects in a system are intimately linked to the nature of its broken symmetry state, which is determined by the specific particle interactions at play. As a result, defects can be seen as unique fingerprints of the system, providing valuable insights into the characteristics of the broken symmetry state. \\
Upon a sudden temperature quench from above $T>T_{BKT}$ to  $T<T_{BKT}$, the initially disordered state of the system becomes unstable, triggering the onset of ordering in the process of the system's evolution towards the low temperature equilibrium state. In this process, defects of opposite signs interact, resulting in their annihilation and the eventual emergence of a homogeneous configuration. As time progresses, the number of defects in the system decreases, ultimately leading to a steady state where only a few defects remain. In two dimensions, the ordering is quasi long-ranged, allowing finite size systems to achieve a defect-free steady state. However, in an infinite size system, this process takes an infinite amount of time, rendering it practically impossible to obtain a defect-free state.

{Defect Detection:} The algorithm used for detection of defects is the following: first, to detect the rotation of the orientation field around each point ($i,j$), we evaluate the discretized version of the integral,
\begin{equation*}
I=\oint_{\Gamma} \vec \nabla \theta.d\vec r =\oint_{\Gamma}d\theta
\end{equation*}
where, $\Gamma$ is a closed contour around the point $(i,j)$ traversed in the counterclockwise sense.\\
This gives, 
\begin{equation*}
\begin{split}
 I=\oint_{\Gamma} \vec \nabla \theta.d\vec r & = 2 k \pi,  \hspace{1mm}if  \hspace{1mm}\Gamma\hspace{1mm} encloses\hspace{1mm} a  \hspace{1mm}defect \hspace{1mm}core\\
 & =0, \hspace{1mm}otherwise
   \end{split}                                         
  \label{eq(2)}
\end{equation*}

Where $k$ is called the winding number, which characterizes the type of defect.  The winding number at lattice point $(i,j)$ is calculated as,
\begin{equation*}
k=\frac{I}{2\pi}
\end{equation*}

In the subsequent sections, the results are derived by considering the loop $\Gamma$ around each point $(i, j)$, encompassing its eight neighboring points. To assess the effect of the discrete nature of integration \cite{canova2016competing, jensen1992phenomenological}: we conducted additional simulations using a larger integration loop. We found that our results remain unaffected. However, it is essential to exercise caution, as employing an excessively large integration loop may lead to incorrect outcomes. \\
For a system with polar symmetry topologically stable defect configurations correspond to $k=\pm1$ whereas for a system with nematic symmetry of spins, topologically stable defect configurations are those with $k=\pm\frac{1}{2}$. The energy cost of the formation of a defect pair of winding number $\pm k$ is $\sim k^2$. Hence, the formation of higher order defects leads to a much larger free energy cost for the system, and at the same time, the higher order defects are not topologically stable in two dimensions. Hence, even though the higher order defects may arise in the system, they are very transient and wiped out very quickly. Therefore, we are interested only in defects with $k=\pm 1$ and $\pm \frac{1}{2}$ in systems with polar and nematic symmetry, respectively.

\twocolumngrid
\begin{figure*}
\centering
\hspace{-0.3 in}\subfloat{\includegraphics[width=0.90\textwidth]{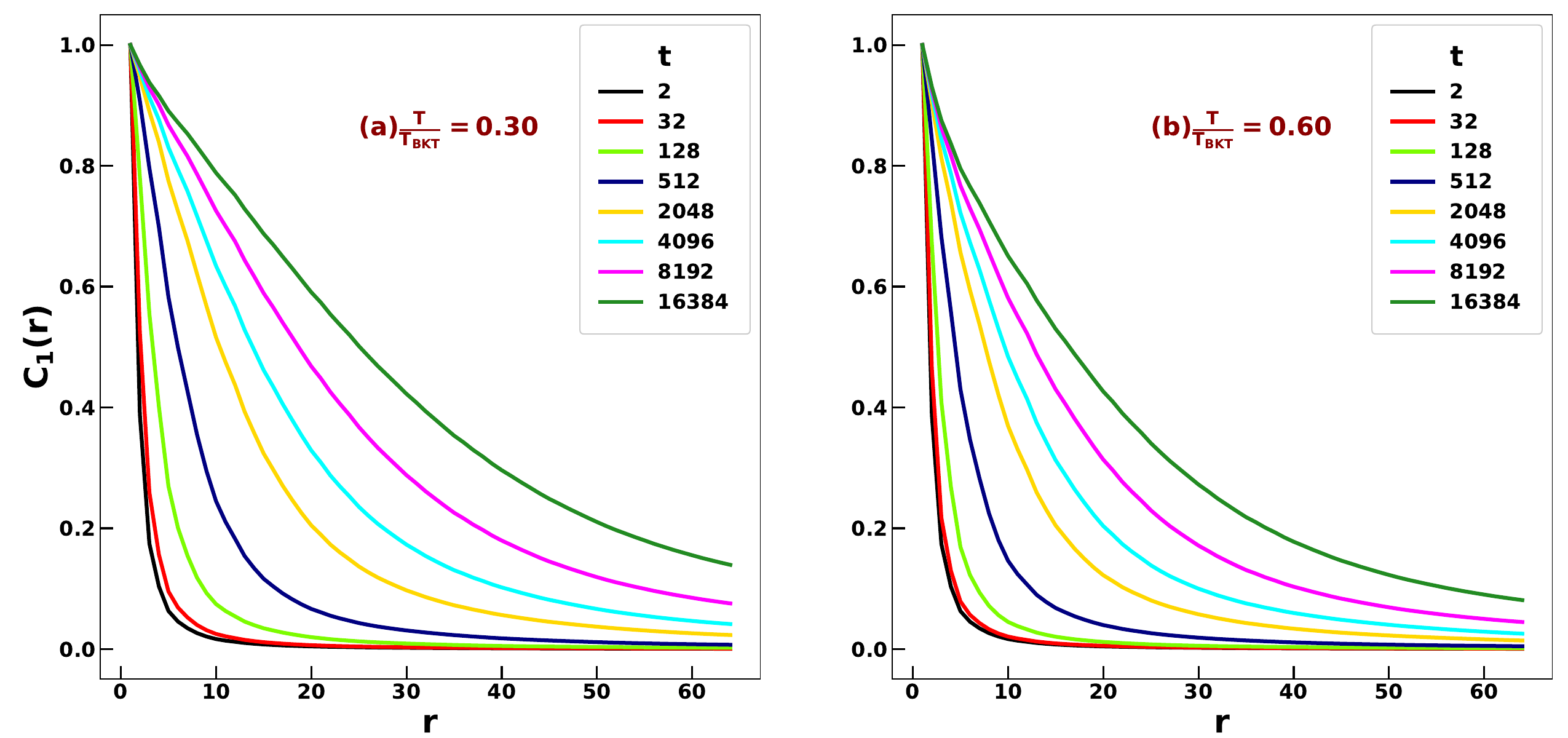}}

\subfloat{\includegraphics[width=0.45\textwidth]{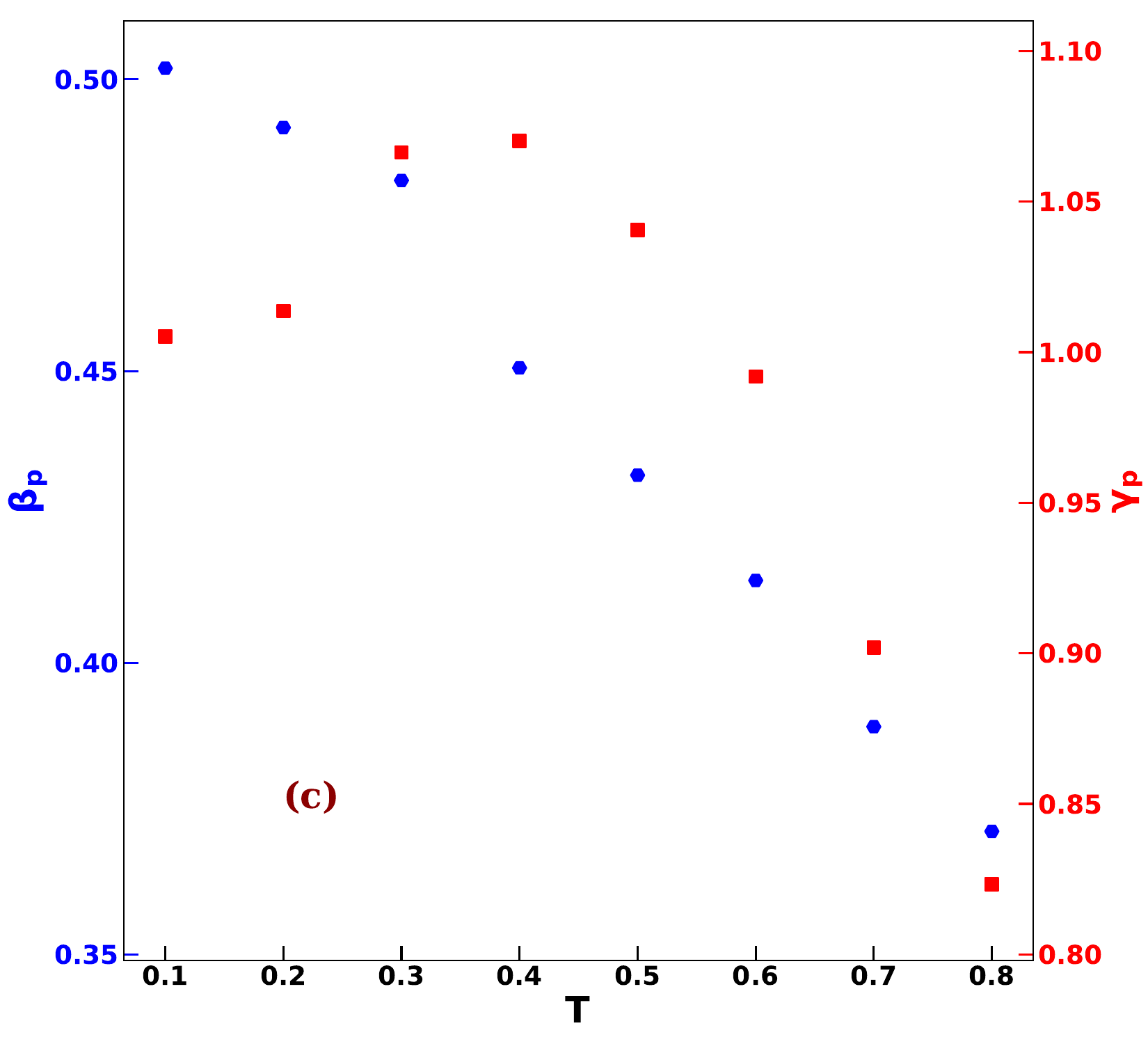}}

\caption{ (color online) The figure illustrates the behavior of the correlation function, $C_1(r)$, for the pure polar system at two different temperatures: (a) $\frac{T}{T_{BKT}} = 0.30$, and (b) $\frac{T}{T_{BKT}} = 0.60$. Subplot (c) depicts the variation of the exponents with temperature for the following: correlation length, $L(t) \sim \bigg(\frac{t}{ln(t)} \bigg)^{\beta_p}$ and number of defects, $N_p(t) ln\{N_p(t)\} \sim t^{-\gamma_p}$; System size $L=128$.}
\label{fig:5}
\end{figure*}

\twocolumngrid
\begin{figure*}
\centering
\hspace{-0.3 in}\subfloat{\includegraphics[width=0.90\textwidth]{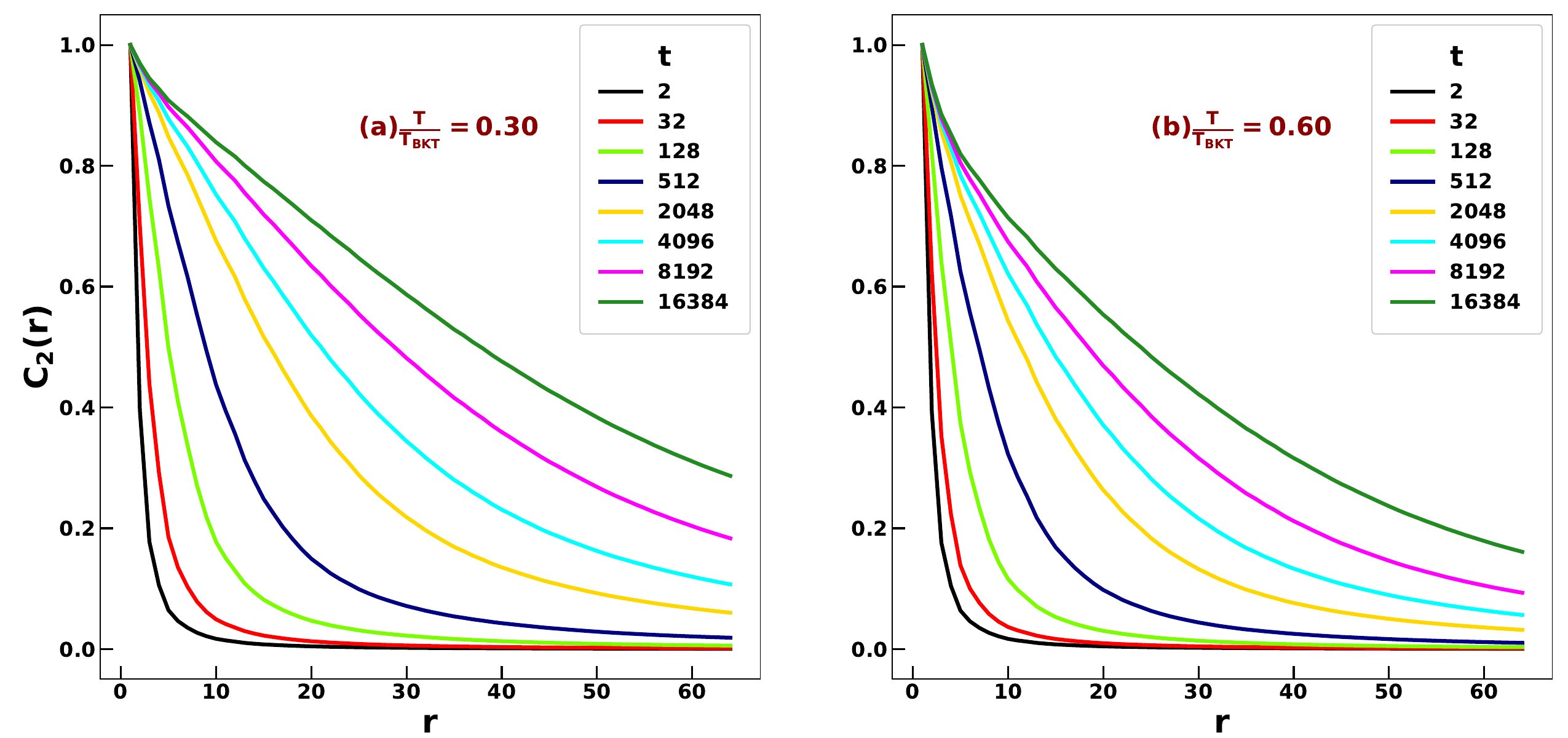}}

\subfloat{\includegraphics[width=0.45\textwidth]{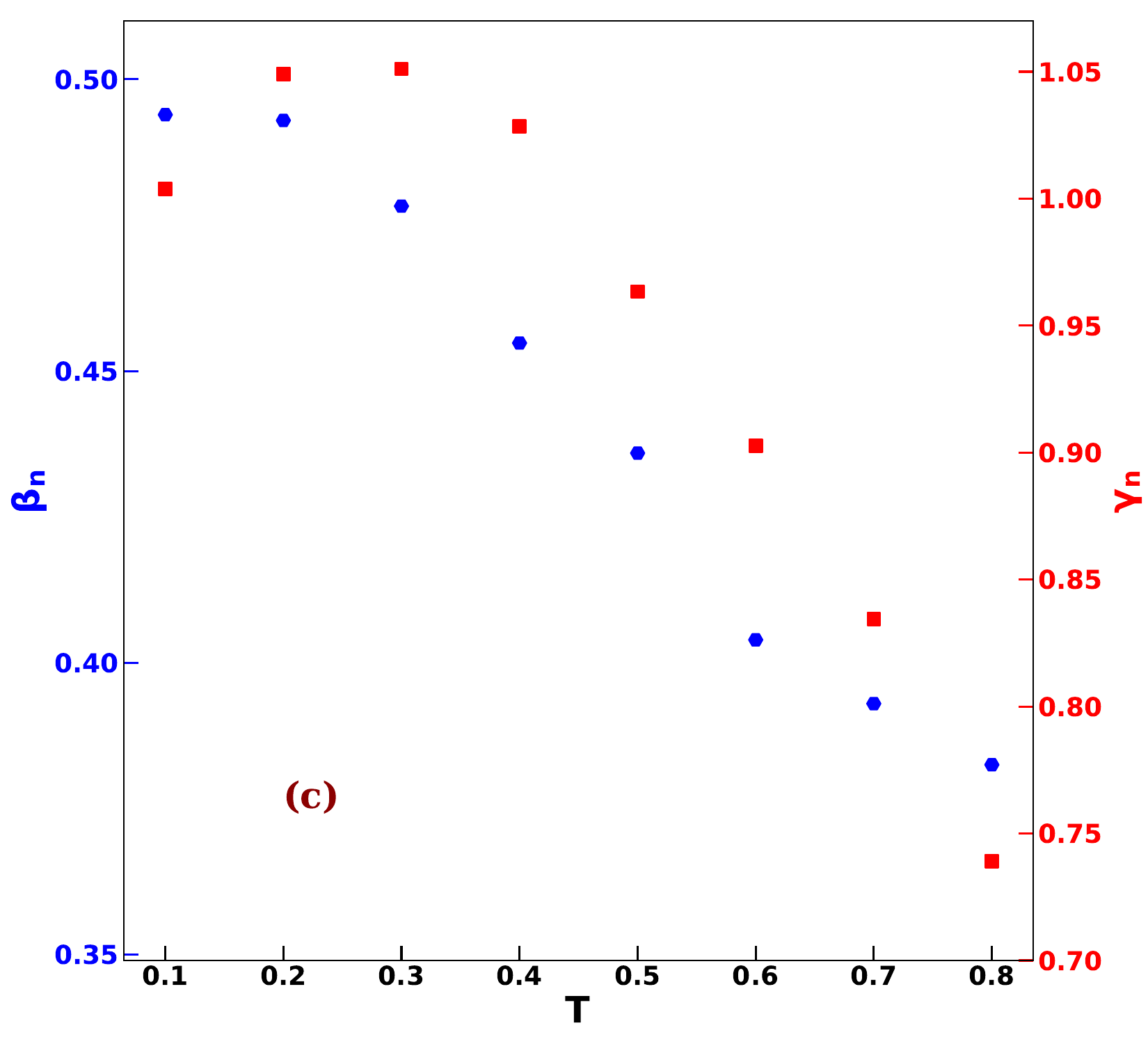}}

\caption{ (color online) The figure illustrates the behavior of the correlation function, $C_2(r)$, for the pure apolar system at two different temperatures: (a) $\frac{T}{T_{BKT}} = 0.30$, and (b) $\frac{T}{T_{BKT}} = 0.60$. Subplot (c) depicts the variation of the exponents with temperature for the following: correlation length, $L(t) \sim \bigg(\frac{t}{ln(t)} \bigg)^{\beta_n}$ and number of defects, $N_n(t) ln\{N_n(t)\} \sim t^{-\gamma_n}$; System size $L=128$.}
\label{fig:6}
\end{figure*}

\section{Results}\label{secIII}
\subsection{Pure Polar System}\label{secIIIA1}

Topologically stable $\pm 1$ defect structures in a system governed solely by ferromagnetic interactions are shown in fig.\ref{fig:1}(a). These defects move through the system exhibiting diffusive behavior \cite{muzny1992direct,HaraP}. Upon collision, oppositely charged defects annihilate, leading to the emergence of a locally homogeneous, uniformly oriented state. The fig.\ref{fig:2} captures the temporal evolution of the system, while the decay in the number of defects over time is illustrated through the $N_p(t)$ $vs.$ $t$ plot at various temperatures, as shown in fig.\ref{fig:3}(a).  In this representation, $N_p(t)$ is defined as, $N_p(t) = \frac{1}{2}\left[\mathlarger{\sum}_{k=+1}^{-1} n_{k}(t)\right]$ where $n_{\pm1}(t)$ denotes the count of $k=\pm1$ defects at time $t$. Thus, $N_p$ represents the average number of integer defects (average of the number of positive and negative defects) in the system at time $t$.
Across all temperatures in the $T<T_{BKT}$ regime, the $N_p(t)$ $vs.$ $t$ plot exhibits a power-law decay characterized by $N \sim t^{-\alpha_p}$, where $\alpha_p$ is the exponent characterising the decay \cite{mynote3}.
Previous studies of the same have established that in a system of spins interacting solely through ferromagnetic interactions, the exponent $\alpha_p$ is approximately 1 with logarithmic correction \cite{Jelić_2011,koo2006,Yurke1993}.
However, in our study, we have observed that $\alpha_p(T)$ exhibits temperature dependence. When the temperature is significantly lower than the critical temperature $T_{BKT}$, $\alpha_p(T)$ maintains a value close to 1, which concurs with previous research findings. But, as the temperature is increased, there is a noticeable reduction in the value of $\alpha_p(T)$ as shown in fig.\ref{fig:3}(c). The distinctive trend in $\alpha_p(T)$'s behavior can be elucidated as follows:\\
The motion of defects in the system is governed by the interplay between two fundamental factors: the cooperative interaction between spins, which promotes alignment, and the thermal energy of the spins, which introduces randomness. At extremely low temperatures ($T \ll T_{BKT}$), the thermal energy is negligible compared to the dominant spin interaction, resulting in the prevalence of well-defined spin waves \cite{maccari2020}. This regime is characterized by a nearly constant value of $\alpha_p$ close to 1 (with logarithmic correction), and the cooperative behavior of spins dominates. However, as the temperature increases, the thermal energy starts to play a more significant role, leading to an increased frequency of random spin flips. Moreover, elevation in temperature induces transverse fluctuations in the well-ordered spin arrangement, leading to a reduction in spin wave stiffness. Consequently, the system becomes more susceptible to a broader spectrum of spin orientation fluctuations in terms of frequency. This alteration in spin wave stiffness results in the localization of defects and  an increased defect count within the system  contributes to a slowing down of the decay rate of the defect population. As a consequence, the $N_p(t)$ $vs.$ $t$ curve depicted in fig.\ref{fig:3}(a) exhibits a flattened profile, and the decay exponent in fig.\ref{fig:3}(c) experiences a decline. These observations signify a transition from a regime dominated by spin waves to one influenced by thermal effects.

\subsection{Pure Nematic System}\label{secIIIA2}
Stable defect configurations in a system of apolar spins, corresponding to $k = \pm \frac{1}{2}$ \cite{de1993physics,schan,priestly2012introduction}, are depicted in fig.\ref{fig:1}(b).
These are known as disinclinations.  The number of defects in this system also exhibits a time-dependent decay, as observed in the evolution shown in fig.\ref{fig:4}. The decay follows a power law behavior characterized by $N_n \sim t^{-\alpha_n}$, where $\alpha_n$ represents the exponent governing the decay \cite{mynote3}. Previous studies have reported an exponent, $\alpha_n$ $\sim 1$ for this decay (with logarithmic correction) \cite{Hickl2021DynamicsOT,chuang1991cosmology,harth2020topological}. However, our study reveals the existence of a temperature dependent exponent $\alpha_n(T)$. In the low-temperature regime dominated by spin waves, the exponent remains close to 1, consistent with previous findings. As the temperature rises, the manifestation of a reduced rate of defect annihilation becomes evident through the flattening of the $N_n(t)$ $vs.$ $t$ curve, depicted in fig.\ref{fig:4}(b). Here, $N_n(t)$ is defined as, 
$N_n(t) = \frac{1}{2}\left[\mathlarger{\sum}_{k=+\frac{1}{2}}^{-\frac{1}{2}} n_{k}(t)\right]$, where $n_{\pm \frac{1}{2}}(t)$ denotes the count of $k = \pm \frac{1}{2}$ defects at time $t$. Thus, $N_n(t)$ represents the average number of $\frac{1}{2}$ integer defects (average of positive and negative defect numbers) in the system at time $t$. Consequently, the exponent of the power law decay, $\alpha_n$, decreases as shown in fig.\ref{fig:4}(d). This behavior can be explained via the same argument as given for the polar case: an increase in temperature causes decay of reduction of spin wave and localization of defects, finally resulting in a slower rate of defect annihilation.\\
In the case of pure systems, both polar and nematic, our results demonstrate a shift in the dominant mechanism governing the system's overall behavior with an increase in temperature—from a region dominated by spin waves to one where thermally influenced effects prevail. This shift in the dominant mechanism contributes to  slower annihilation  of defects, consequently yielding a temperature-dependent exponent. To visually illustrate this deceleration of the annihilation of defects, a series of snapshots and animations are presented in our supplementary material. \\

To assess the influence of the finite size of the system on the exponent's behavior, we conducted simulations for system sizes $L =$ 128, 256, and 512. Notably, the exponents $\alpha_{m,p}$ exhibited a consistent trend across these different system sizes.
Furthermore, the deceleration in the annihilation  of defects suggests a more sluggish evolution of the system towards its equilibrium, characterised by slower growth in correlation among the spins. To delve into this phenomenon, we calculate the equal time two point spatial correlation of spin orientations in the system denoted by $C_1(r,t)$ and $C_2(r,t)$ in the polar and nematic case, respectively. Mathematically, these correlations are defined as, $C_1(r,t) = \bigg<cos(\theta(r_0+r,t)-\theta(r,t))  \bigg>$ and $C_2(r,t) = \bigg<cos(2\theta(r_0+r,t)-2\theta(r,t))  \bigg>$ , where, $<...>$  signify an average over the reference point $r_0$ and multiple independent realizations. The temporal evolution of $C_1(r,t)$ and $C_2(r,t)$ at two different temperatures are displayed in FIG.\ref{fig:5}(a-b) and FIG.\ref{fig:6}(a-b), respectively, in the supplementary material. The increasing correlation over time indicates the progressive growth of domains. The correlation lengths for the pure polar and nematic cases, denoted as $L_p(t)$ and $L_n(t)$, are defined at the 0.5 crossing of $C_1(r,t)$ and $C_2(r,t)$, respectively. These lengths, encapsulated within $L_{p,n}(t)$, represent the characteristic size of domains in the system at time $t$. In a 2D system with continuous spin symmetry, represented by a vector order parameter,  $L_{p,n}(t)$ is predicted to vary as: $L_{p,n}(t) \sim (\frac{t}{ln(t)})^{\beta_{p,n}}$, where $\beta_{p,n}$ denotes the exponent for the pure polar and pure nematic systems, respectively. The established value for these exponents is $\frac{1}{2}$ \cite{bray2002theory,bray2000breakdown}.\\
Further, to incorporate the impact of logarithmic corrections in the decay of the number of defects, we examine the behavior of $N(t)ln(N(t))$, which is known to behave as $N(t)ln(N(t)) \sim t^{-\gamma}$, where, $N(t) \equiv$ $N_p(t)$ and $N_n(t)$ for pure polar and pure nematic cases, respectively, with corresponding exponents $\gamma_p$ and $\gamma_n$.  The previously established value of $\gamma$ is 1\cite{Yurke1993}. However, our observations reveal that at low temperatures both $\beta_{p,n}$ and $\gamma_{p,n}$ closely match with previously observed values. However, with an increase in temperature, $\beta_{p,n}$ and $\gamma_{p,n}$ exhibit a decay, as illustrated in FIG.\ref{fig:5}(c) and FIG.\ref{fig:6}(c) for the pure polar and pure nematic systems, respectively.\\
The diminishing values of $\beta_{p,n}$ with increasing temperature suggest a gradual slowdown in domain growth for both pure polar and pure nematic cases. This deceleration can be attributed to the slower annihilation  of defects.
\\

Next, we present our findings for the mixed system, where both types of interactions are concurrently at play.

\subsection{Mixed System}\label{secIIIA3}
In the mixed system, every spin concurrently interacts with its neighbors through both ferromagnetic and nematic pathways, with the relative strength of the two interactions determined by $\Delta$. The intricate interplay between these two types of interactions can result in intriguing system behavior within the intermediate range characterized by $0 < \Delta < 1$, bridging the gap between the pure polar and pure nematic phases observed in the limiting cases. In this section, our primary exploration focuses on understanding the system's behavior near the two extreme limits. We investigate how the system responds when deviating slightly from the pure polar or pure nematic states and explore how this response correlates with the magnitude of the deviation. \\

Before presenting our results, it is important to define the observables for the mixed system. For $\Delta$ values close to 1, we showcase the results for the number of integer defects, denoted as $N_p(t)$ in the previous section. Additionally, in this range, we examine the behavior of the correlation function $C_1(r,t)$ and corresponding correlation length $L_{mp}$ varying as:  $L_{mp}\sim (\frac{t}{ln(t)})^{\beta_{mp}}$, where $\beta_{mp}$ denotes the exponent. On the other hand, for $\Delta$ values close to 0, we present the results for the number of half-integer defects, previously denoted as $N_n(t)$. In this range, we also investigate the behavior of the correlation function $C_2(r,t)$ and the corresponding correlation length $L_{mn}$ varying as:  $L_{mn}\sim (\frac{t}{ln(t)})^{\beta_{mn}}$, where $\beta_{mp}$ denotes the exponent.\\

\begin{figure} [hbt]
{\includegraphics[width=1.0 \linewidth]{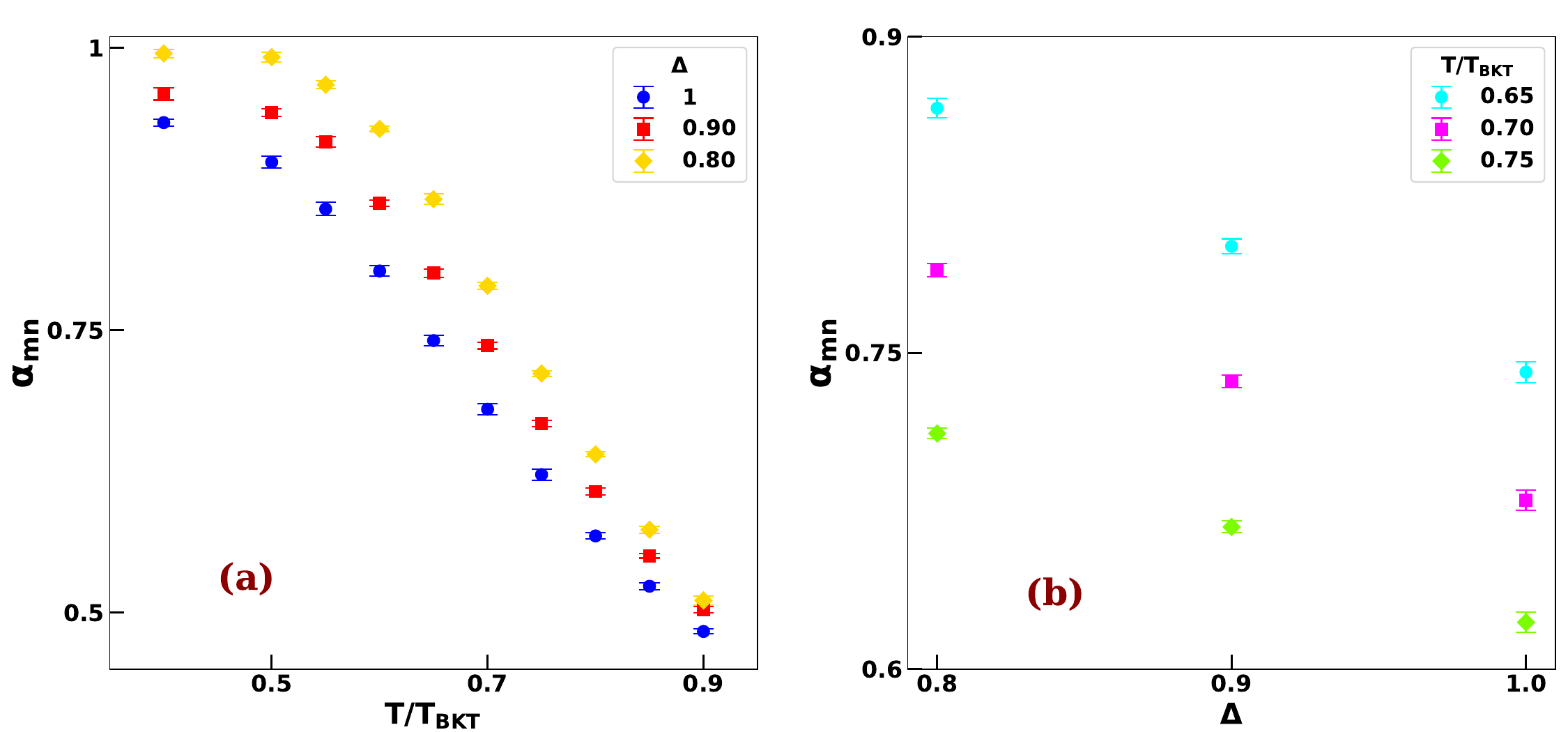}}
	\caption{(color online) The figure showcases the variation of the exponent $\alpha_{mp}$ within the Polar Region with error bars representing the uncertainty in the exponent value at each data point. Panel (a) displays the exponent's behavior with temperature for various $\Delta$ values, while panel (b) depicts its variation with $\Delta$ across different temperatures. The remaining parameters remain consistent with those presented in fig.\ref{fig:3}.}
\label{fig:7}
\end{figure}

\begin{figure} [hbt]
{\includegraphics[width=1.0 \linewidth]{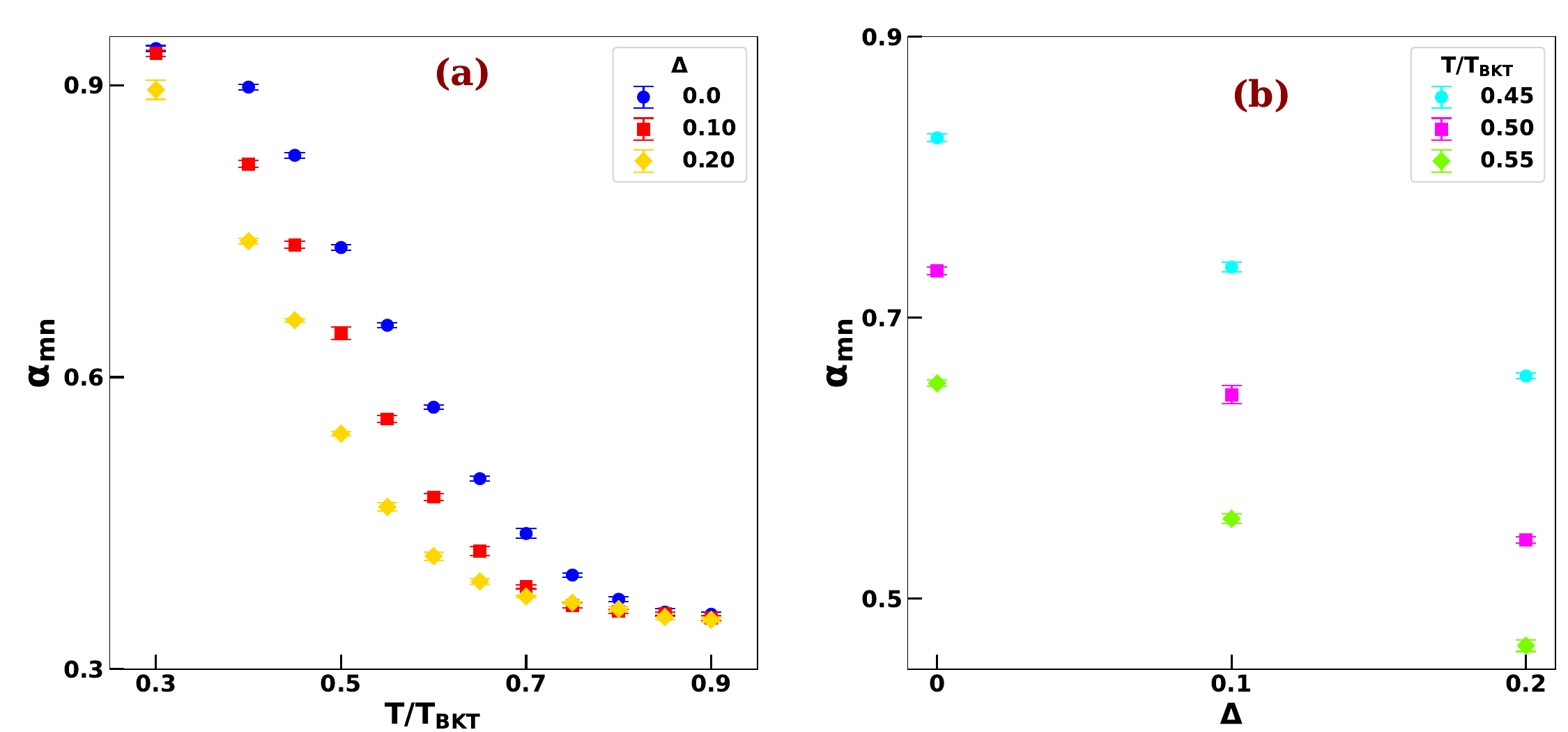}}
	\caption{(color online) The depicted figure illustrates the change in the exponent $\alpha_{mn}$ within the Nematic Region with error bars representing the uncertainty in the exponent value at each data point. Panel (a) demonstrates the exponent's temperature dependence for different $\Delta$ values, while panel (b) illustrates its variation with $\Delta$ for different temperatures. The rest of the parameters remain the same as depicted in fig.\ref{fig:3}.}.
\label{fig:8}
\end{figure}

\twocolumngrid
\begin{figure*}
\centering
\hspace{-0.2 in}\large{\color{red}{\bf Polar Region}}\hspace{0.34\linewidth}\large{\color{red}{\bf Nematic Region}}\\
\hspace{-0.3in}\large{\color{blue}${\bf \Delta = 0.8}$}\hspace{0.17\linewidth}\large{\color{blue}${\bf \Delta = 0.9}$}\hspace{0.19\linewidth}\large{\color{blue}${\bf \Delta = 0.1}$}\hspace{0.17\linewidth}\large{\color{blue}${\bf \Delta = 0.2}$}
\vspace{0.05 in}
\hspace{-0.6 in}\subfloat{\includegraphics[width=0.52\textwidth]{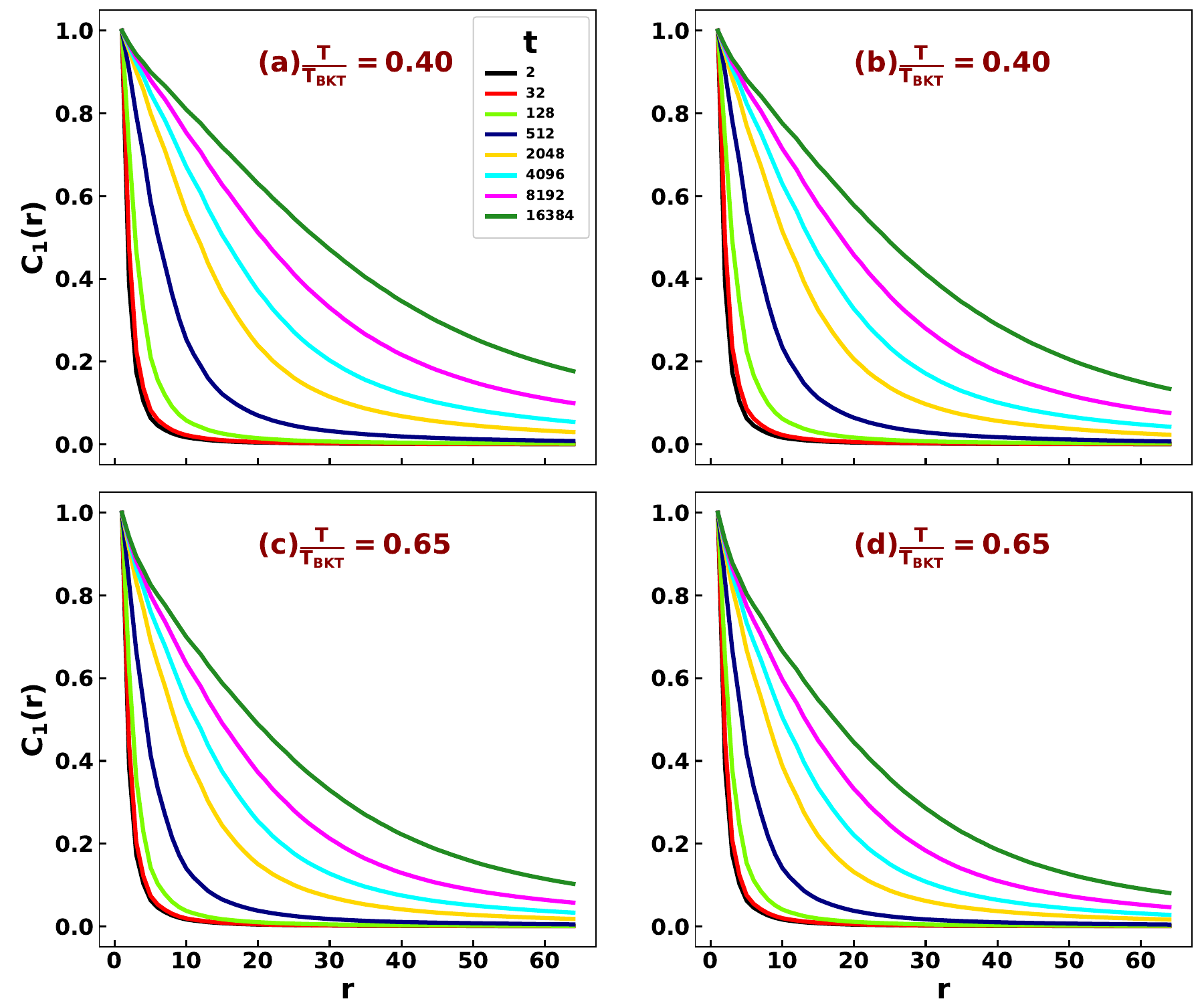}}
~
\subfloat{\includegraphics[width=0.52\textwidth]{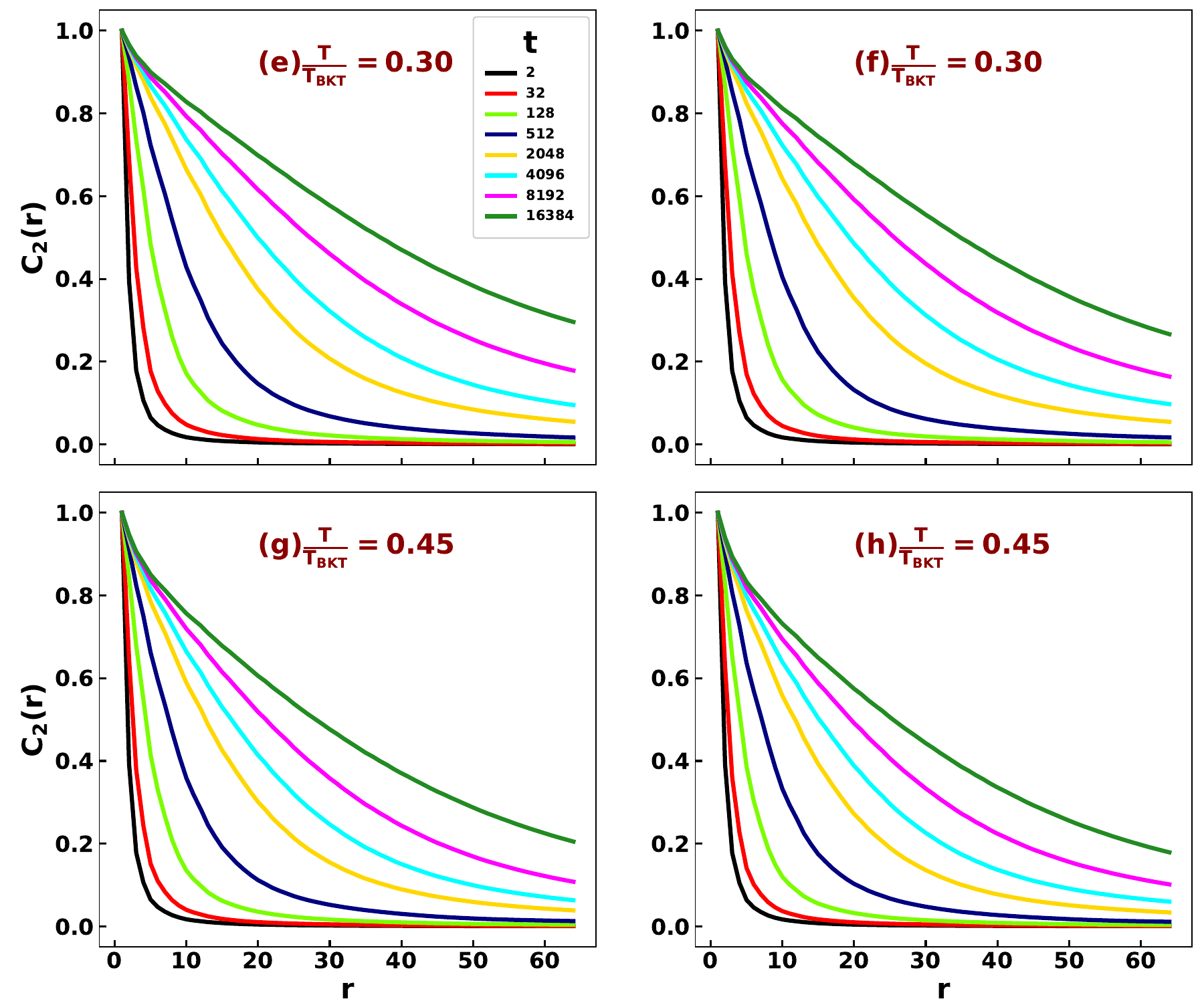}}

\hspace{-0.3 in}\subfloat{\includegraphics[width=0.70\textwidth]{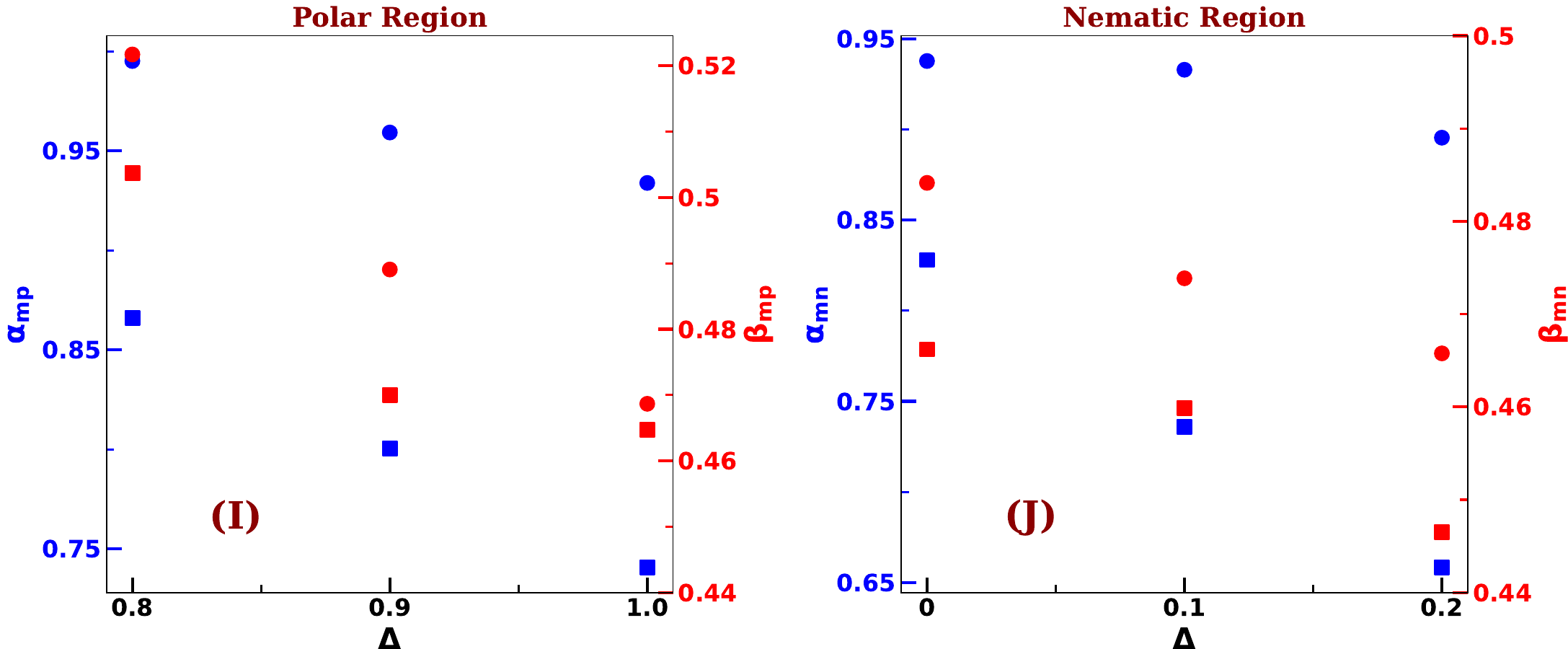}}

\caption{(color online) The figure illustrates the behavior of the correlation function for the mixed system in the polar region (a-d) and the apolar region (e-h). In the polar region, we show the results for the correlation function, $C_1(r)$, two values of $\Delta =$ 0.9, 0.8 at two different temperatures : (a-b) $\frac{T}{T_{BKT}}=0.40$, and (c-d) $\frac{T}{T_{BKT}}=0.65$. In the apolar region, we show the results for the correlation function, $C_2(r)$, for two values of $\Delta =$ 0.1, 0.2 at two different temperatures : (e-f) $\frac{T}{T_{BKT}}=0.30$, and (g-h) $\frac{T}{T_{BKT}}=0.45$. The correlation length, L(t), in both the polar and apolr region follows the behavior: $L(t) \sim \bigg(\frac{t}{ln(t)} \bigg)^{\beta}$, where $L(t) = L_{mp}(t) \& \beta = \beta_{mp}$ for the polar region and $L(t) = L_{mn}(t) \& \beta = \beta_{mn}$ for the nematic region; (I) Depicts the variation of exponents $\alpha_{mp}$ and $\beta_{mp}$ in blue and red, respectively, in the polar region. Different symbols \(\scalebox{2}{\(\bullet\)}\) and \(\scalebox{1.2}{\(\blacksquare\)}\) are used for two different temperatures $T= 0.40$ and $0.65$, respectively.; (J) Depicts the variation of exponents $\alpha_{mn}$ and $\beta_{mn}$ in blue and red, respectively, in the nematic region. Different symbols \(\scalebox{2}{\(\bullet\)}\) and \(\scalebox{1.2}{\(\blacksquare\)}\) are used for two different temperatures $T= 0.30$ and $0.45$, respectively.; System size $L=128$.}
\label{fig:9}
\end{figure*}

In proximity to the limit of pure polar limit, we examined three values of $\Delta$ = ($0.8$, $0.9$, and $1.0$). Within this range, the system exhibited the prevalence of $\pm 1$ defects. The temporal decay of the $\pm 1$ defect population followed a power-law pattern, expressed as $N(t) \sim t^{-\alpha_{mp}(\Delta)}$, where $\alpha_{mp}(\Delta)$ denotes the exponent governing the decay rate. The temporal evolution of the defect population exhibited a power-law decay, described by $N(t) \sim t^{-\alpha_{mp}(\Delta)}$, where $\alpha_{mp}(\Delta)$ represents the exponent governing the decay rate. In the low-temperature regime, the exponent $\alpha_{mp}(\Delta)$ closely approached $1$ for all values of $\Delta$. However, with an increase in temperature, the exponent progressively decreased, as depicted in fig. \ref{fig:7}(a). The reduction in the magnitude of $\alpha_{mp}(\Delta)$ can be  attributed to the shift in mechanism  controlling the system's behavior from a region dominated by spin waves to one dominated by temperature-induced effects, akin to the pure cases. Additionally, a decrease in $\Delta$ from $1.0$ to $0.8$ at a fixed temperature revealed a prominent upward trend in $\alpha_{mp}$, as illustrated in fig. \ref{fig:7}(b).\\
The increasing trend of  $\alpha_{mp}$ with increase in $\Delta$ can be explained as follows: 
Decreasing $\Delta$ from 1.0 brings nematic interactions into play. It is well-known that creating a pair of integer defects incurs a significantly higher energy cost compared to a pair of half-integer defects. Consequently, the system tends to favor the formation of a few half-integer defects at the expense of integer defects. This results in a reduced number of integer defects and a faster rate of annihilation for them, leading to an increased $\alpha_{mp}$ at higher temperatures, as depicted in fig. \ref{fig:7}(a). However, at lower temperatures, where the system contains only a few defect pairs, the polar alignment tendency far outweighs the nematic alignment tendency. Consequently, altering the value of $\Delta$ has minimal impact on the number of defect pairs. As a consequence, $\alpha_{mp}$ remains nearly constant and close to 1 for all values of $\Delta$, primarily due to the dominant influence of spin waves.

Approaching the nematic boundary, we investigated three distinct values of $\Delta$ = ($0.2$, $0.1$, and $0$). In this region, the system predominantly exhibits $\pm \frac{1}{2}$ defects. The number of $\pm \frac{1}{2}$ defects follows a power-law decay: $N \sim t^{-\alpha_{mn}}$, where the exponent $\alpha_{mn}$ remains close to 1 at low temperatures. As the temperature increases, $\alpha_{mn}$ decreases, as illustrated in fig. \ref{fig:8}(a), consistent with the behavior observed in the pure nematic case. However, on increasing $\Delta$ from 0 at a fixed temperature, $\alpha_{mn}$ exhibits a downward trend, as shown in fig. \ref{fig:8}(b).
The observed behavior can be attributed to the dominance of nematic interaction among neighboring spins at a finite but lower $\Delta$ values. Moreover, the lower energy required to create a half-integer defect leads to the formation of a large  number of $\pm \frac{1}{2}$ defects at the same temperature.  As a result, the annihilation of defects with opposite signs occurs more rapidly, leading to an increased value of $\alpha_{mn}$, as illustrated in fig.\ref{fig:8}(b).\\

Moreover, to characterize the impact of  $\Delta$ variation on domain growth in both the polar and nematic regions, we analyze the behavior of the correlation functions and the corresponding exponents of the correlation lengths, as detailed at the beginning of this section. The corresponding results at two distinct temperatures are illustrated in FIG.\ref{fig:9}. In the polar region, there is an increase in $\beta_{mp}$ as $\Delta$ decreases from 1. Conversely, in the nematic region, the exponent  $\beta_{mn}$ decreases with an increase in $\Delta$ from zero. Thus, the trends observed for $\beta_{mp}$ and $\beta_{mn}$ align seamlessly with those of $\alpha_{mp}$ and $\alpha_{mn}$ in the polar and nematic regions, respectively.

Overall, these findings demonstrate the intricate interplay between spin interactions, temperature, and number of defects in the mixed system, unveiling fascinating behavior in the intermediate regime of $\Delta$ values near the pure polar and  pure nematic limit.

\twocolumngrid
\begin{figure*} [t]
\centering
\subfloat[]{\includegraphics[width=0.42\textwidth]{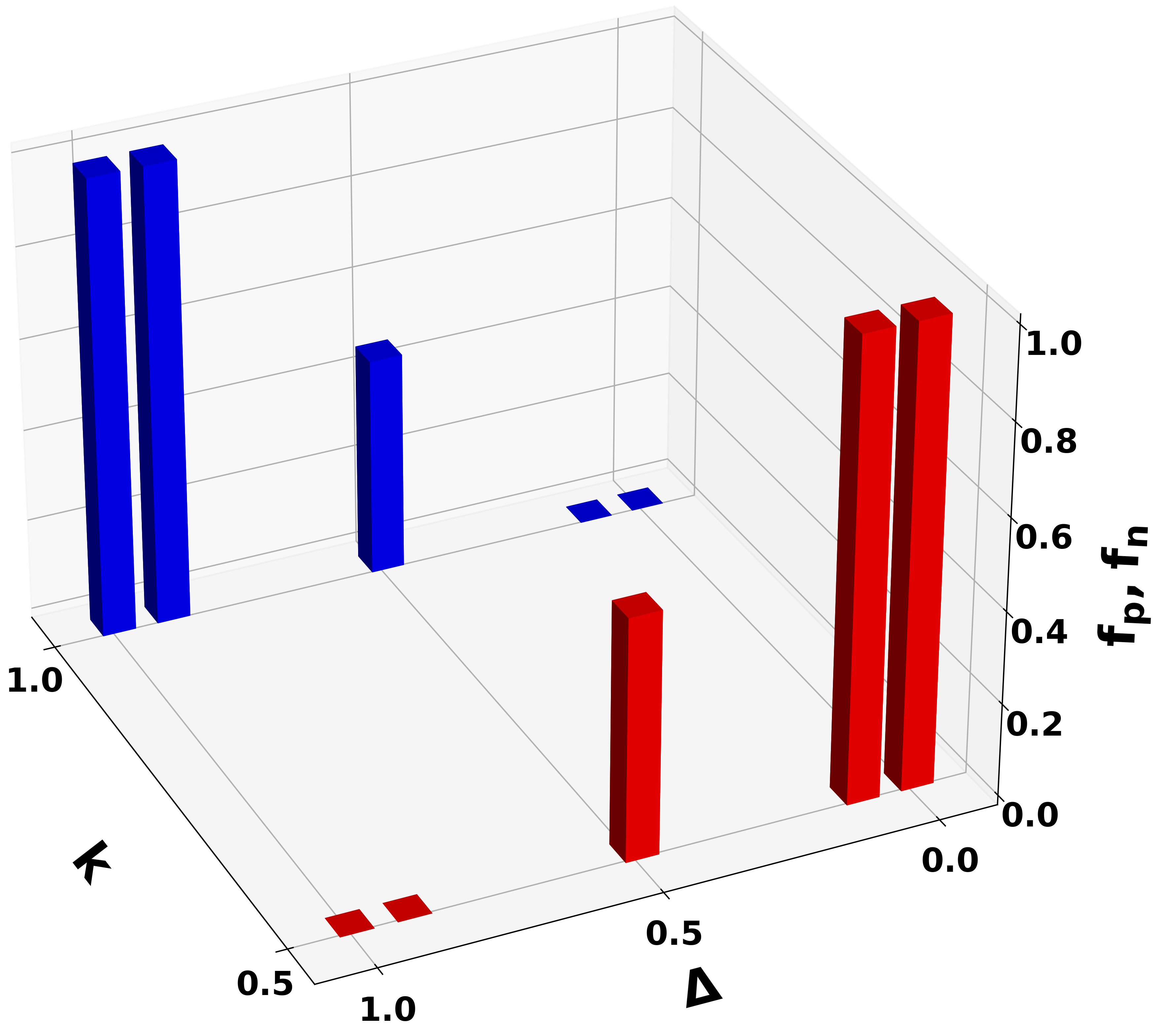}}
~
\subfloat[]{\includegraphics[width=0.42\textwidth]{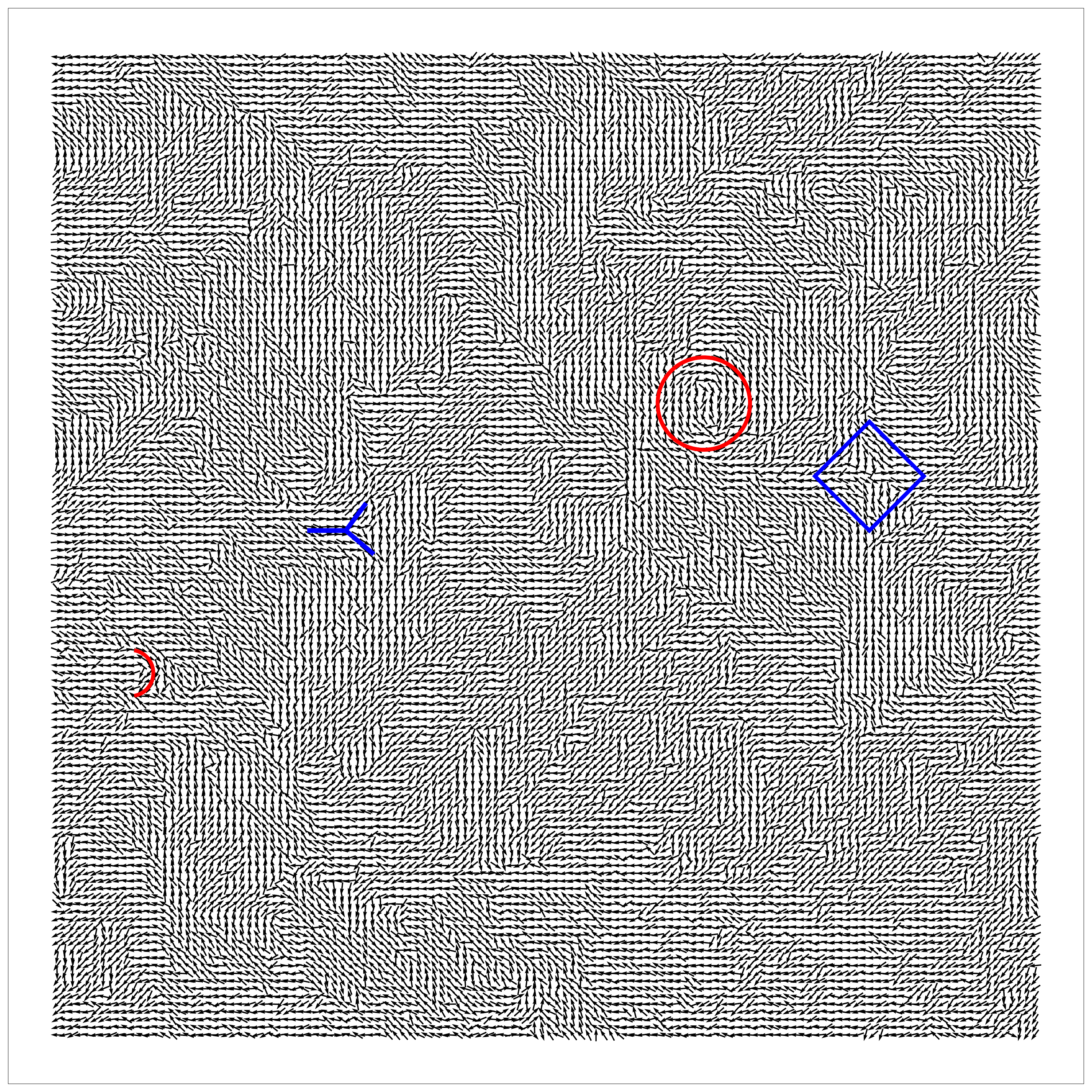}}  

\caption{(color online)(a) The figure depicts the fractions, $f_p$ and $f_n$, of $\pm 1$ and $\pm \frac{1}{2}$ defects out of total number of defects within distinct phases at a temperature of $\frac{T}{T_{BKT}} = 0.60$ at time t=2000, with the 'blue' color representing $f_p$ and 'red' color indicating $f_n$. All other parameters remain consistent with those presented in fig.\ref{fig:3}; (b) The figure portrays the configuration of the system in a coexistence state at $\frac{T}{T_{BKT}} = 0.6$ and $\Delta = 0.50$ at time t = 4000. In this visualization, spins are represented by black quivers.  A pair of $\pm 1$ defects is denoted by a red circle and blue square, while a pair of $\pm \frac{1}{2}$ defects is indicated by a red arc and a blue 'Y'-shaped structure. The remaining parameters are consistent with fig.\ref{fig:2}.}
\label{fig:10}
\end{figure*}

\twocolumngrid
\begin{figure*} [hbt]
\subfloat[]{\includegraphics[width=0.495\textwidth]{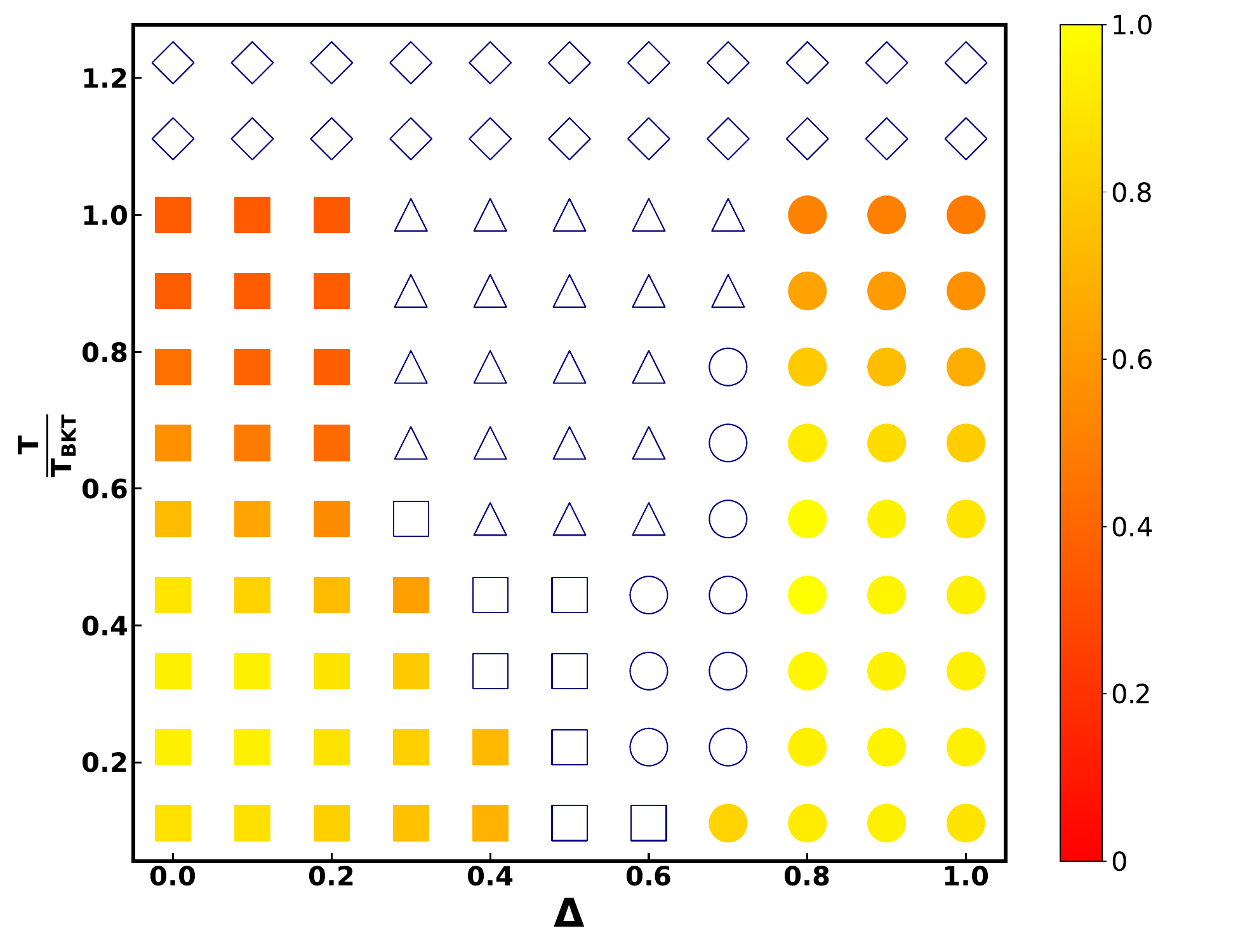}}
~
\subfloat[]{\includegraphics[width=0.495\textwidth]{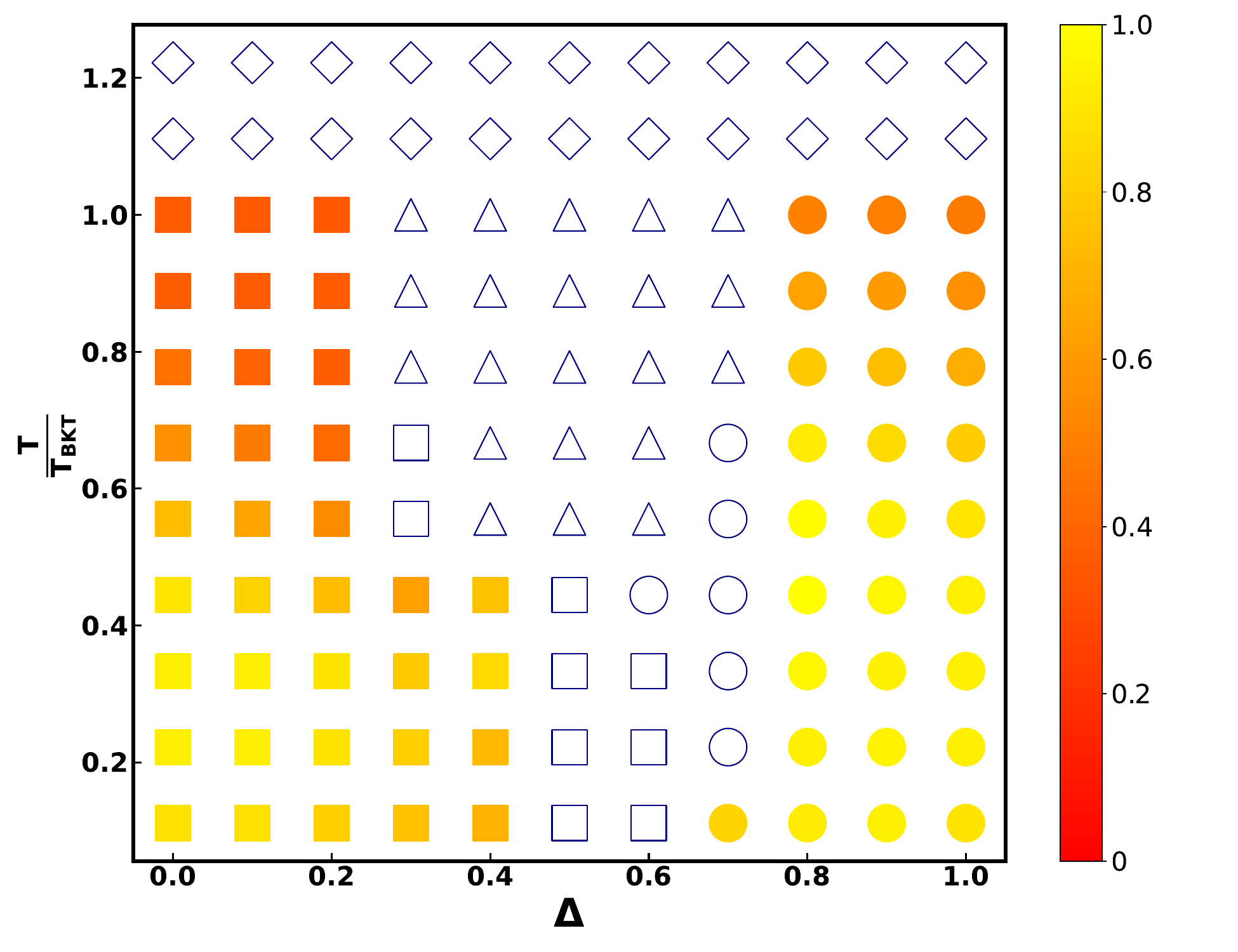}}
\caption{(color online) This diagram illustrates the Phase Diagram of the system, where distinct symbols are utilized to depict different phases, and their colors correspond to the associated exponent values as indicated by the color bar. Specifically, the symbols `circle', `square', `triangle', and `diamond' represent the Polar, Nematic, Coexistence, and Disordered phases, respectively. Within the Polar and Nematic regions, the filled symbols indicate sub-regions where the exponent can be calculated, while the unfilled symbols signify parameter values where the exponent calculation is not feasible. To illustrate the dependence of phases on system size, we present the phase diagram for two different system sizes: (a) $L = 128$ and (b) $L = 256$.  All other parameters remain consistent with those in fig.\ref{fig:3}. }
\label{fig:11}
\end{figure*}

\begin{figure} [hbt]
\subfloat[]{\includegraphics[width=0.495\textwidth]{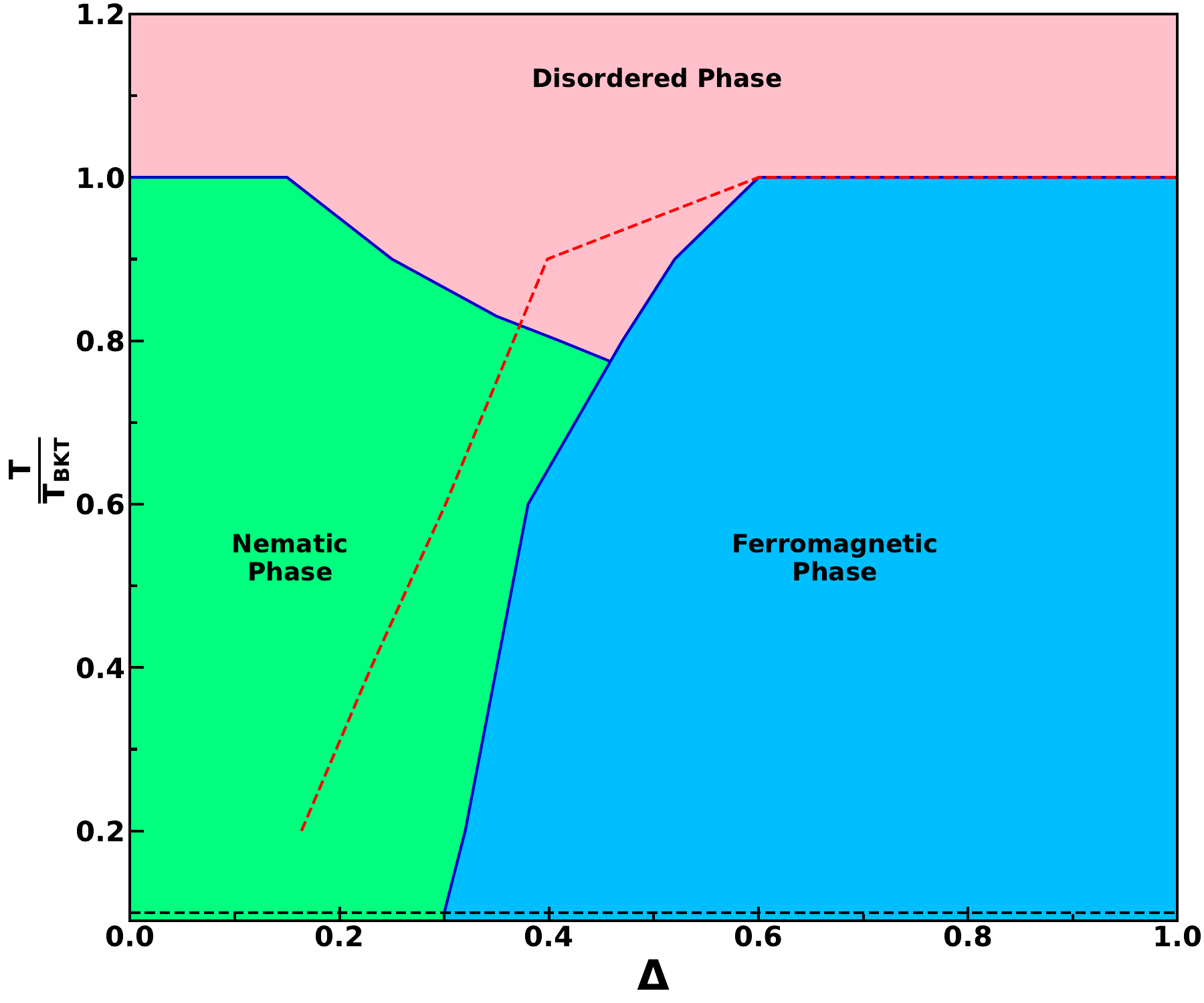}}
\caption{(color online)  The figure illustrates equilibrium phase diagram obtained in our model. The boundary between the different phases are marked by `blue' solid line for system size $L = 128$. Dashed line in `red' represents the phase boundary obtained on extrapolating the data for system sizes $L =$ $64$, $128$, $160$ and $200$. The `black' horizontal dashed line at the bottom marks the line for $\frac{T}{T_{BKT}} = 0.1$
}

\label{fig:12}
\end{figure}

\subsection{Kinetics Phase Diagram}\label{secIIID}

In the previous section, we investigated the behavior of defects in pure systems ($\Delta = 0$ and $\Delta = 1$) as well as in the mixed system ($0<\Delta<1$) near the boundaries of pure polar and pure nematic interactions. By analyzing the exponent of the power law decay of the defect population, we discerned the impact of temperature in the pure systems and the effect of varying $\Delta$ in the mixed system.

Continuing with our analysis, we delve into the characterization of various phases within the system by investigating the behavior of defects across the entire parameter space of $\Delta$ and $T$. Our observations indicate the existence of three distinct phases below $T_{BKT}$ : Polar Phase, Nematic Phase and Coexistence Phase, alongside the emergence of a disordered phase for temperatures above $T_{BKT}$. These phases can be explained as follows:

{\bf Polar Phase : } In this phase, the system demonstrates the emergence of integer defects, which signifies the dominance of ferromagnetic interactions and, consequently, the prevalence of integer defects. When the temperature remains below $T_{BKT}$ and $\Delta$ is nearly equal to 1, the count of integer defects follows a power-law decay over time. A comprehensive understanding of the exponent's behavior is outlined in fig. \ref{fig:10}. Nevertheless, as the value of $\Delta$ experiences a slight increase, despite integer defects retaining their dominance, the characteristic decay pattern slightly deviates from the typical power-law behavior. As a result, calculating an exponent within this specific region becomes unattainable.

{\bf Nematic Phase:} This phase is distinguished by the exclusive prevalence of half-integer defects. For temperatures below $T_{BKT}$ and $\Delta$ approaching 0, the count of half-integer defects adheres to a power-law decay. The specifics of the exponent's behavior are expounded upon in fig. \ref{fig:11}. However, with a slight increment in $\Delta$, the power-law decay pattern dissipates, even though the system predominantly consists of half-integer defects. Consequently, determining the exponent becomes infeasible.

{\bf Coexistence Phase:} In the vicinity of $\Delta = \frac{1}{2}$ with a slightly elevated temperature (commencing from well below $T_{BKT}$), where the strengths of both types of interactions are comparable, the system possesses ample thermal energy. This results in a competitive interplay among these interactions, leading to a unique spin configuration wherein both polar and nematic phases coexist.

{\bf Disordered Region:} This phase occurs when the temperature exceeds $T_{BKT}$. The thermal energy becomes dominant and overwhelms both types of interactions, resulting in a disordered arrangement of spins.

To substantiate our aforementioned categorization, we calculate the fractions $f_p$ and $f_n$ corresponding to integer and half-integer defects, respectively. These fractions are defined as follows: $f_p = \frac{<n_I>}{<n_I>+<n_{HI}>}$ and $f_n = \frac{<n_{HI}>}{<n_I>+<n_{HI}>}$. $<n_I>$ and $<n_{HI}>$ represent the average numbers of integer and half-integer defects in the system at a given time $t$, considering a specific set of parameters $(T, \Delta)$, where $<....>$ denotes average over 100 independent realizations. \\
In fig.\ref{fig:10}(a), we present the bar graph  of $f_p$ and $f_n$ in various regions, as described above, at temperature $\frac{T}{T_{BKT}} = 0.60$ at time $t=2000$. When $\Delta = 1$ or $0.9$, the values $f_p \approx 1$ and $f_n \approx 0$ indicate the exclusive formation of $\pm 1$ defects in the system. Conversely, for $\Delta = 0$ or $0.1$, the values $f_p \approx 0$ and $f_n \approx 1$ suggest the exclusive presence of $\pm \frac{1}{2}$ defects in the system. Therefore, as $\Delta$ approaches 1, a polar phase is evident, while values of $\Delta$ close to 0 signify a nematic phase. Notably, at $\Delta = 0.50$, both $\pm 1$ and $\pm \frac{1}{2}$ defects coexist in the system, resulting in nearly equal values of $f_p$ and $f_n$, indicating  almost equal probability for the formation of both types of defects. This region is referred to as the coexistence region.\\
To provide empirical evidence supporting the coexistence of both types of defects at $\Delta = 0.5$, we present a snapshot of the system's configuration in fig. \ref{fig:10}(b).  Within this snapshot, spins are visually represented through black quivers. The visual representation showcases the co-occurrence of integer and half-integer defects. Specifically marked within the snapshot is a pair of integer and half-integer defects for reference.

The fig.\ref{fig:11} displays the complete phase diagram of our system, where each symbol corresponds to a specific phase based on the given parameter values : `circle', `square', `triangle', and `diamond' represent the Polar, Nematic, Coexistence, and Disordered phases, respectively. Filled symbols indicate that the exponent of the power law decay has been calculated and the color of the symbol represents the value of the exponent as indicated by the color bar. On the other hand, empty symbols represent parameter values for which the exponent could not be calculated. This comprehensive phase diagram provides a visual representation of the different phases and their corresponding exponent values in our system.

We identify various regimes in the $(T,\Delta)$ space based on the nature (integer or half-integer) of the topological defects for two different system sizes ($L = 128 \& 256$). The boundary of the different phases remains almost the same for both system sizes. Importantly, we avoid the characterization of phase transitions. Instead, we observe that within a specific parameter range where polar and nematic interaction strengths are comparable, both types of defects coexist in the system. Our emphasis on characterizing domain growth is specifically concentrated in the vicinity of the pure polar and pure apolar limits, where domain growth is notably influenced by variations in the value of $\Delta$, as detailed in the preceding section.\\

\textbf{Equilibrium Phase Diagram:}
In Figure \ref{fig:12}, we present the equilibrium phase diagram derived from our model, revealing three distinct phases previously identified in literature: Ferromagnetic (F), Nematic Phase (N), and Disordered Phase (D). The classification of these phases relies on two correlation functions, namely $C_1(r,t)$ and $C_2(r,t)$ as discussed in Section \ref{secIIIA3} and based on their behavior we draw boundary between different phases. The `blue' solid line in Fig.\ref{fig:12} denotes the phase boundaries for a system size of $L=128$. Notably, the boundaries between F $\to$ D and N $\to$ D align well with previous findings. However, the boundary between F $\to$ N, while exhibiting some similarity, deviates noticeably from established results. We attribute this discrepancy to a system size effect. To substantiate our argument, we performed extrapolations for the boundary between the F $\to$ N phases using data from system sizes $L =$ $64$, $128$, $160$, and $200$. The extrapolated phase boundary is shown with red 'dashed' lines in Fig. \ref{fig:12}.\\

 We can also compare the equilibrium phase diagram with the kinetic phase diagram obtained from the kinetics of the system as shown in Fig.\ref{fig:11} and Fig.\ref{fig:12}. For  $\Delta$ close to 1 and $\Delta$ close to zero and near $T_{BKT}$ (order-disorder temperature) both phase diagrams match well. It can be observed that the coexistence phase in our kinetic phase diagram appears for the parameter range close to the Nematic to Ferromagnetic transition line.

 {\section{Discussion}\label{secdis}}
 In this study, we investigate the attributes of a spin system situated on a 2D square lattice, where each spin has the flexibility to align within the range of $[0, 2\pi]$. These spins engage with their immediate neighbors through ferromagnetic and nematic interactions simultaneously with the relative strength of the two determined by the parameter $\Delta$. Employing the Markov chain Monte Carlo algorithm, we systematically explore the $(T, \Delta)$ space to gain insights into defect behaviors.\\
 In the context of the generalized XY model, the preceding studies have outlined a typical phase diagram featuring two low-temperature phases (ferromagnetic (F) and nematic (N)), along with a high-temperature disordered phase (D). The characterization of these phases primarily relied on the behavior of two-point correlation functions  \cite{D_B_Carpenter_1989,hubscher2013stiffness,nui2018correlation}, susceptibility \cite{dian2011spin}  and the transition between phases was marked by the behavior of the magnetization order parameter, as well as the features of the binder cumulant \cite{canova2014kosterlitz}, stiffness jump across the transition\cite{hubscher2013stiffness,canova2014kosterlitz}, behavior of specific heat \cite{QI2013127}. The transitions $F \to D$ and $N \to D$ were categorized as BKT type \cite{hubscher2013stiffness,canova2014kosterlitz}
, while the transition $F \to N$  is an Ising transition \cite{QI2013127}.\\
However, our present work explores the annihilation  of defects and its influence on domain growth during the transient time regime, as the system progresses towards a steady state following the initial quench.\\
Our examination uncovers distinctive findings in different regions of the $(T, \Delta)$ plane, particularly within the context of mixed systems ($0 < \Delta < 1$). 
In the scenario of pure systems, where the parameter $\Delta$ takes on values of either 0 or 1, a distinct pattern emerges. As the temperature increases within the moderate to high range, without surpassing the critical temperature $T_{BKT}$, the exponent governing the power-law decay of defect count experiences a reduction. This observation indicates a noteworthy trend: the rate at which defects are removed becomes slower in this temperature range. In contrast, at lower temperatures, the exponent remains close to 1. This results in a significant transition as temperature changes. With increasing temperature, the system shifts from being predominantly influenced by the spin waves to fluctuations due to thermal noise. In essence, there is a marked change in the dominant mechanisms that dictate the system's behavior.

Furthermore, when examining mixed system—fascinating and intricate behaviors come to the forefront.
For temperatures exceeding $T_{BKT}$, disorder prevails regardless of the $\Delta$ value. However, for temperatures below $T_{BKT}$, we discern three distinct regions: In proximity to the polar limit ($\Delta$ values close to 1), the system primarily exhibits dominance of integer defects. Near the nematic limit (approaching $\Delta = 0$), half-integer defects prevail. Around $\Delta \approx \frac{1}{2}$, an intriguing coexistence region manifests, characterized by the simultaneous presence of integer and half-integer defects. This intriguing phase has not been previously documented in earlier studies as per our knowledge. Further, the equilibrium phase diagram obtained in our study shows very good match with those obtained in previous studies. \\
These findings shed light on the intricate interplay between temperature and interaction parameters, revealing diverse defect behaviors within the spin system. The observed decrease in exponent values in pure cases and the identification of the emergence of a novel phase in mixed scenarios add a layer of depth to our understanding of the system's behavior.\\
 The system explored in our study can be realised in liquid crystals \cite{lee1985strings,domany1984first,pang1992string} and superfluid $^3$He films \cite{korshunov1985possible,bhaseen2012discrete}. Existing experimental setups, originally devised for the pure XY model \cite{nagaya1992experimental} and the `twisted nematic liquid crystal' \cite{almeida2021phase} offer the capability to detect variations in defect numbers, spin correlations, and other relevant factors. The versatile nature of the experimental design presented in  \cite{almeida2021phase} suggests the feasibility of employing a similar methodology to validate the findings of our study.

\section{Acknowledgement}
P.S.M., P.K.M. and S.M., thanks PARAM Shivay for computational facility under the National Supercomputing Mission, Government of India at the Indian Institute of Technology, Varanasi and the computational facility at I.I.T. (BHU) Varanasi. P.S.M. and P.K.M.  thank UGC for research fellowship. S.M. thanks DST, SERB (INDIA), Project No.: CRG/2021/006945, MTR/2021/000438  for financial support. 

\section{Data Availability}
The data that support the findings of this study are available within the article.

\bibliography{citation}

\begin{thebibliography}{58}
\bibitem [{myn()}]{mynote3}%
  \BibitemOpen
  \href@noop {} {\bibinfo  {journal} {We observe that at high temperatures ($T<T_{BKT}$), the power law behavior is not followed at late times. However, at intermediate times (the region marked in the fig.3(a-b)) we obtained a good power law fit.}\ }\BibitemShut {NoStop}%
\end{thebibliography}

\end{document}